# Interstellar glycolamide: A comprehensive rotational study and an astronomical search in Sgr  B2(N)

M. Sanz-Novo[1], A. Belloche[2], J. L. Alonso[1], L. Kolesniková[1, 3], R. T. Garrod[4], S. Mata[1],
H. S. P. Müller[5], K. M. Menten[2], Y. Gong[2]

[1] Grupo de Espectroscopía Molecular (GEM), Edificio Quifima, Área de Química-Física, Laboratorios de Espectroscopía y Bioespectroscopía, Parque Científico UVa, Unidad Asociada CSIC, E-47011 Valladolid, Spain. jlalonso@qf.uva.es

[2] Max-Planck-Institut für Radioastronomie, Auf dem Hügel 69, 53121 Bonn, Germany

[3] Department of Analytical Chemistry, University of Chemistry and Technology, Technická 5, 166 28 Prague 6, Czech Republic

[4] Departments of Chemistry and Astronomy, University of Virginia, Charlottesville, VA 22904, USA

[5] I. Physikalisches Institut, Universität zu Köln, Zülpicher Str. 77, 50937 Köln, Germany

June 24, 2020

## ABSTRACT

*Context.* Glycolamide is a glycine isomer and also one of the simplest derivatives of acetamide (e.g., one hydrogen atom is replaced with a hydroxyl group), which is a known interstellar molecule.

*Aims.* In this context, the aim of our work is to provide direct experimental frequencies of    the ground vibrational state of glycolamide in the centimeter-, millimeter- and submillimeter-wavelength regions in order to enable its identification in the interstellar medium.

*Methods.* We employed a battery of state-of-the-art rotational spectroscopic techniques in the frequency and time domain to measure the frequencies of glycolamide. We used the spectral line survey named Exploring Molecular Complexity with ALMA (EMoCA), which was performed toward the star forming region Sgr B2(N) with ALMA to search for glycolamide in space. We also searched for glycolamide toward Sgr B2(N) with the Effelsberg radio telescope. The astronomical spectra were analyzed under the local thermodynamic equilibrium approximation. We used the gas-grain chemical kinetics model MAGICKAL to interpret the results of the astronomical observations.

*Results.* About 1500 transitions have been newly assigned up to 460 GHz to the most stable conformer, and a precise set of spectroscopic constants was determined. Spectral features of glycolamide were then searched for in the prominent hot molecular core Sgr B2(N2). We report the nondetection of glycolamide toward this source with an abundance at least six and five times lower than that of acetamide and glycolaldehyde, respectively. Our astrochemical model suggests that glycolamide may be present in this source at a level just below the upper limit, which was derived from the EMoCA survey. We could also not detect the molecule in the region's extended molecular envelope, which was probed with the Effelsberg telescope. We find an upper limit to its column density that is similar to the column densities obtained earlier for acetamide and glycolaldehyde with the Green Bank Telescope.

## 1. Introduction

N-bearing molecules have attracted the attention of much theoretical and experimental research in the last half-century. Only a few molecules containing the $NH_2$ group have been confidently detected in the interstellar medium (ISM). However, chemical models predict the formation of complex interstellar amines, such as methoxyamine ($CH_3ONH_2$) and N-methylhydroxylamine ($CH_3NHOH$). Recently, we have chemically liberated both amines (Kolesniková et al. 2018, 2017) from their corresponding hydrochloride salts; additionally, precise sets of spectroscopic constants were obtained and used to search for spectral features of both molecules in the molecular clouds Orion KL, Sgr B2, B1-b, and TMC-1.

Moreover, it is interesting to study molecules that contain a peptide bond given that this type of bond plays an essential role as a building block of proteins, which are fundamental components of living systems (Belloche et al. 2017). The simplest molecule containing a peptide bond, formamide ($NH_2CHO$), was detected in the interstellar medium (ISM) in the 1970s in Sagittarius B2 (Rubin et al. 1971) and later in Orion KL (Turner 1989, 1991). One of the simplest derivatives of formamide that also contains a peptide bond, acetamide ($CH_3C(O)NH_2$), has already been detected in the ISM (Hollis et al. 2006; Halfen et al. 2011). Following this, we propose the study of glycolamide, an amide derivative ($CH_2(OH)C(O)NH_2$), as a candidate for interstellar detection.

Glycolamide is an interesting compound for various reasons. On the one hand, it was detected experimentally by Nuevo et al. (2010) when they simulated the relevant conditions to the interstellar medium and Solar System icy bodies such as comets. In their experimental setup, a condensed $CH_3OH : NH_3 = 11 : 1$ ice mixture was UV irradiated at about 80 K and urea, glycolic acid, glycerol, glycerolic acid, and glycerol amide were also found as products. Also, reactions involving glycolamide are relevant, since it can participate in polycondensation reactions, leading to the for-mation of N-(2-amino-2-oxoethyl)-2-hydroxyacetamide, a derivative of glycine dipeptide (Szőri et al. 2011), which was already investigated by high-resolution rotational spectroscopy (Cabezas et al. 2017).

Of paramount importance is the fact that glycolamide is also an isomer of glycine, the smallest amino acid. Attempts to observe glycine in the ISM have been reported but its detection has never been confirmed (Snyder et al. 2005; Cunningham et al. 2007; Jones et al. 2007; Hollis et al. 2003). Nevertheless, several amino acids have already been found in some chondritic meteorites (Pizzarello 2006; Pizzarello et al. 2010; Burton et al 2012). Recently, Altwegg et al. (2016) have reported the presence of volatile glycine together with the precursor molecules methylamine and ethylamine in the coma of comet 67P/Churyumov-Gerasimenko. Furthermore, the observation of not only amino acids but also their most essential isomers as plausible precursors, in the ISM and in solar system bodies, would be of crucial importance for revealing the chemistry that may have led to life's origin (Ehrenfreund et al. 2001). It is known that total dipole-moment values can be used as a discrimination element for the detection of rotational transitions. An inspection of the most essential glycine isomers shows that the dipole moment of glycolamide is double that the one of methylcarbamate and N-methyl carbamic acid, and even a factor four higher than that of glycine conformer I (Sanz-Novo et al. 2019; Lattelais et al. 2011). Thus, glycolamide should be a good candidate to search for in the ISM. Among all the isomers studied up to now, methylcarbamate has been the only one searched for in the hot molecular cloud W51e2, associated with high-mass protostar, and the intermediate-mass protostar IRAS21391+58502 (Demyk et al. 2004). Groner et al. (2007) studied the millimeter and submillimeter-wave spectrum of methyl carbamate, providing rotational data in wide frequency ranges. Finally, glycolamide has been characterized by Stark modulation spectroscopy between 60 and 78.3 GHz (Maris 2004).

In this work, we use a battery of state of the art rotational spectroscopic techniques in the frequency and time domain to derive a precise set of spectroscopic constants for glycolamide. The $^{14}$N nuclear quadrupole coupling constants need to be determined together with the rotational constants in order to correctly reproduce the spectrum, which is a prerequisite for its identification in astronomical spectra. This is especially true for observations in low-frequency regions accessible with the Green Bank Telescope (GBT) or the Effelsberg 100 m telescope, for which the hyperfine components of transitions with low angular momentum number J are generally spread over several MHz. Moreover, glycolamide stands as a challenging problem for gas-phase millimeter-wave rotational studies due to its high melting point (122–126 °CC), which is related to condensation problems. Therefore, its spectrum so far has remained unexplored in these high-frequency regions. High-quality data at mm and submm wavelengths are also necessary for the eventual astronomical detection of glycolamide with observational facilities such as the radio telescopes of the Institut de RadioAstronomie Millimétrique (IRAM) or the Atacama Large Millimeter-submillimeter Array (ALMA). This motivated us to record the millimeter and submillimeter-wave spectrum of glyco-lamide up to 460 GHz using the heated-cell millimeter-wave spectrometer at the University of Valladolid.

One of the best astronomical sources to search for complex organic molecules in the interstellar medium is the giant molecular cloud complex Sagittarius B2 (Sgr B2) that is located in the galactic center region at a distance of 8.2 kpc (Reid et al. 2019). This star forming region contains the protocluster Sgr B2(N) in which several hot molecular cores are embedded (e.g., Bonfand et al. 2019). Many complex organic molecules were first detected toward Sgr B2(N) (see, e.g., McGuire 2018, for a census). Several spectral line surveys were done toward Sgr B2(N) over the past four decades. We use one of the most recent ones, the imaging spectral line survey EMoCA, which stands for Exploring Molecular Complexity with ALMA (Belloche et al. 2016), performed with ALMA, to search for glycolamide and compare its abundance to other related organic molecules. We also targeted the whole Sgr B2(N) region with the Effelsberg telescope to search for glycolamide in its extended molecular envelope.

We describe in Sect. 2 the experimental setup that was used to measure the spectrum of glycolamide. Section 3 presents the results of our analysis of the experimental data. The results of a search for glycolamide toward the hot molecular core Sgr B2(N2) are reported in Sect. 4 along with a comparison to other related molecules. Section 5 reports on our search for glycolamide toward Sgr B2(N) with the Effelsberg telescope. The astronomical results are then put into a broader astrochemical context by comparing them to predictions of numerical simulations in Sect. 6. The conclusions of this work are given in Sect. 7.

## 2. Experimental details

A commercial sample of glycolamide was purchased from Aldrich and used without further purification (98 %). It is a white crystalline substance melting at 122–126 °C. Rotational spectra measurements were carried out with three different spectroscopic techniques.

### 2.1. CP-FTMW spectroscopy

The supersonic-jet rotational spectrum in the 9.2–17.1 GHz frequency range was investigated using a broadband CP-FTMW spectrometer (Mata et al. 2012; Kolesniková et al. 2017). A pulsed heated nozzle was used to expand a gas mixture containing the gaseous glycolamide (heated from the solid at 130 °CC) seeded in neon (backing pressure of 1 bar). Up to 60,000 individual free induction decays (four FIDs on each valve cycle) were averaged in the time domain, and Fourier-transformed using a Kaiser–Bessel window function to obtain the frequency-domain spectrum. The uncertainty of the frequency measurements was estimated to be smaller than 10 kHz.

### 2.2. MB-FTMW and DR-MB-FTMW spectroscopy

In an attempt to fully resolve the $^{14}$N hyperfine structure of glycolamide, we took advantage of the higher sensitivity and resolution of our molecular beam Fourier transform microwave (MB-FTMW) spectrometer, which operates in the frequency range from 4 to 14 GHz and is described elsewhere (Balle & Flygare 1981; Alonso et al. 1997). Glycolamide was conventionally heated at 130 °C and then seeded in Ne (stagnation pressure 1 bar) and expanded adiabatically to form a supersonic jet into a Fabry–Pérot resonator. A microwave radiation pulse sequence that allows for multiple FID collection was recently implemented (León et al. 2017). The microwave transient FID was again registered in the time domain and Fourier transformed into the frequency domain. The pulsed molecular beam was introduced parallel to the axis of the resonator, so each observed transition appears as a Doppler doublet. The resonance frequency is determined by the arithmetic mean of the two Doppler components.

In a next step, we configured a double-resonance (DR-MB-FTMW) technique in our laboratory to extend the coverage of the microwave spectrum of the most stable conformer of glycolamide up to 40 GHz. In this technique (Mori et al. 2009), millimeter-wave radiation was sent into the Fabry-Pérot cavity of the FTMW spectrometer while monitoring a microwave transition by the FTMW system, and transitions induced by the millimeter-wave radiation, even with the hyperfine structure resolved, were observed as a change in the intensity of the monitored transition. This new configuration is of great importance because it allows us to measure in the same frequency region covered by astronomical surveys conducted at cm wavelengths with the GBT, and also serves as a straightforward connection between the microwave and the millimeter-wave data.

### 2.3. MMW spectroscopy

Afterwards, the rotational spectrum was measured from 80 to 460 GHz using the millimeter-wave absorption spectrometer at the University of Valladolid coupled with a new heating cell. A small amount of solid (0.8 g) was introduced in a single neck round bottom flask connected directly to the newly built free space cell of the spectrometer (a 140 cm-long Pyrex cell with an inner diameter of 10 cm and Teflon Windows covered by a heating aluminum blanket) (Alonso et al. 2019). The flask was evacuated ($1 \times 10^{-2}$ mbar) and then was slowly heated using a heating tape controlled by a thermocouple to a temperature of 70 °C. The optical path length of the spectrometer was doubled using a rooftop mirror and a polarization grid. The sample was continuously flowed through the cell (which was heated at 100 °C during the whole experiment), and an optimum gas pressure of about 15 µbar was maintained that allowed us to record the spectra in the millimeter and submillimeter-wave region from 80 to 460 GHz. This spectrometer is based on cascaded multiplication of the output of an Agilent E8257D microwave synthesizer (up to 20 GHz) by a set of active and passive multipliers. In this experiment, to cover the recorded spectral range, amplifier-multiplier chains WR10.0, and WR9.0 (VDI, Inc) in combination with an additional frequency doubler and tripler (WR2.2, WR4.3, WR2.8, VDI, Inc.) were employed. The output of the synthesizer was frequency modulated at $f = 10.2$ kHz with modulation depth between 20 and 40 kHz. After the second pass through the cell, the signal was detected using solid-state zero-bias detectors and was then sent to a phase-sensitive lock-in amplifier with 2f demodulation (time constant 30 ms), resulting in a second derivative line shape. A detailed description of the experimental setup has been given elsewhere (Daly et al. 2014; Alonso et al. 2016). Rotational transitions were measured using an average of two scans: one recorded in increasing and the other in decreasing frequency, and also employing a Gaussian profile function with the AABS package (Kisiel et al. 2005). The experimental uncertainty of the isolated symmetric lines is estimated to be around 50 kHz (higher than the usual 30 kHz due to the corresponding temperature broadening).

## 3. Experimental results and discussion

### 3.1. Jet-cooled rotational spectrum

The CP-FTMW rotational spectrum of glycolamide is shown in Fig. 1a. It appears dominated by intense water dimer $(H_2O)_2$ signals and other lines belonging to known decomposition products that were identified and discarded from the analysis. At the first stage of the line assignment, predictions based on the spectroscopic constants from Maris (2004) were used. Initial inspection of the broadband spectrum was directed to search for a set of intense R-branch $\mu_a$-type rotational transitions (with $J' \leftarrow J'' = 2 \leftarrow 1$) arising from $syn$-glycolamide predicted with significant $\mu_a$ electric dipole-moment component. The initial assignment was extended to other intense $b$-type transitions. After an iterative procedure of fittings and predictions, a total of six rotational transitions were measured. They were least-squares fitted to a rigid rotor Hamiltonian (Pickett 1991). After excluding these lines from the spectral analysis, another weaker progression of $^aR$-branch transitions was discovered and assigned to $anti$-glycolamide.

Rotational transitions exhibit the typical hyperfine structure arising from the interaction of the electric quadrupole moment of the $^{14}N$ ($I = 1$) nucleus with the electric field gradient created at the site of the quadrupolar nucleus by the rest of the electronic and nuclear charges of the molecule. As shown in Fig. 1b, the resolution attained with our broadband CP-FTMW technique was not sufficient to fully resolve the hyperfine patterns. At this point, we took advantage of the high resolution of our narrowband MB-FTMW spectrometer. The analysis began with the measurement of a total of 5 hyperfine components by interpreting the quadrupole coupling pattern for the $2_{02} \leftarrow 1_{01}$ rotational transition (see Fig. 1c) and then extended to other $a$- and $b$- type $R$-branch transitions. Once the analysis of $syn$-glycolamide was completed, new $^bR$ -type transitions belonging to $anti$-glycolamide were identified.

Contrary to $syn$-glycolamide, $anti$-glycolamide exhibits a weak $b$-type rotational spectrum and we were only able to measure $b$-type lines during the MB-FTMW experiment. After iterative fittings and predictions, a set of low-$J$ rotational transitions along with a selection of well-resolved nuclear hyperfine components from the previous Stark study (Maris 2004) was adequately weighted and fitted (Pickett 1991) to a semirigid-rotor Hamiltonian in the $A$-reduction and the $I^r$ representation $H_R^{(A)}$ (Watson 1977), supplemented with a term to take into account the quadrupole interaction $H_Q$ (Foley 1947; Robinson & Cornwell 1953), namely $H = H_R^{(A)} + H_Q$. The Hamiltonian was set up in the coupled basis set $I + J = F$, so the energy levels involved in each transition are labeled with the quantum numbers $J$, $K_a$, $K_c$, and $F$. The final set of rotational constants and nuclear quadrupole coupling constants are listed in the second column of Table 1. In a quest to extend the measurements to higher frequencies and allow for an eventual direct comparison with our Effelsberg observations or the GBT line surveys, we used a double-resonance technique with which also hyperfine components can be resolved. An example of the double resonance spectrum is shown in Fig. 2, where the *b*-type transition, $2_{12} \leftarrow 2_{02}$ is monitored and five hyperfine components of the *a*-type transition $3_{03} \leftarrow 2_{02}$, which is connected to the lower level of the millimeter-wave transition, have been observed. Subsequently, we performed a fit (Pickett 1991) including our microwave and double resonance (DR) lines as well as a selection of transitions from Maris (2004). We used the same Hamiltonian mentioned above. The derived spectroscopic parameters for *syn*-glycolamide are listed in the first column of Table 1. Also, in Table A.1 of the Appendix we provide a comparison between the experimental and **ab initio** predicted quadrupole coupling constants.

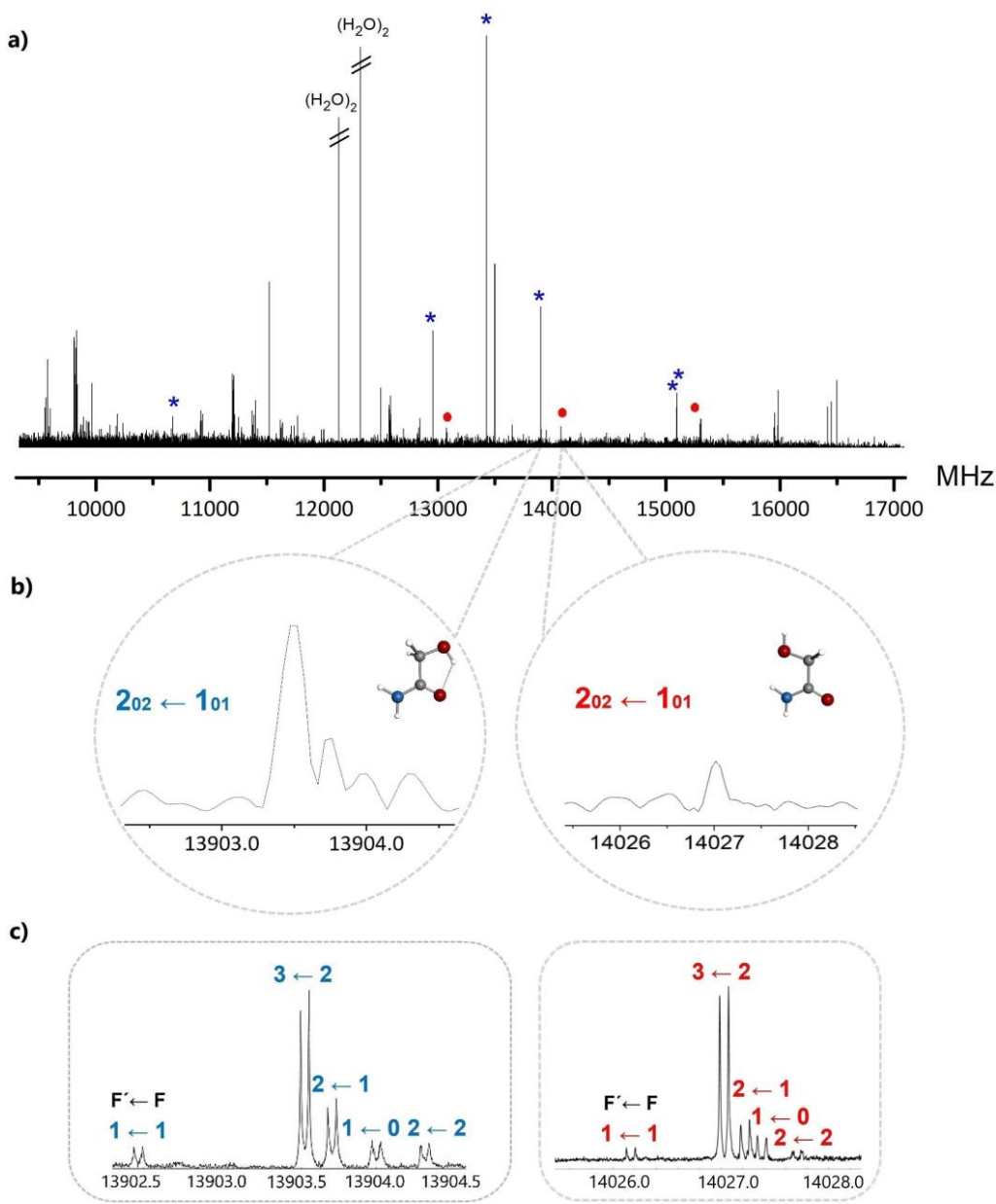

**Fig. 1.** (a) Broadband CP-FTMW spectrum of glycolamide from 9.2 to 17.1 GHz. We depict strong *a*- and *b*-type *R*-branch transitions of *syn*-glycolamide (in blue) as well as some transitions of *anti*-glycolamide (in red). (b) Zoom-in section of the broadband spectrum showing the rotational transition $2_{02} \leftarrow 1_{01}$ of both *syn*-glycolamide and *anti*-glycolamide. (c) The same rotational transition measured in the MB-FTMW experiment, displaying the high resolution of our spectrometer. The resonance frequency is determined by the arithmetic mean of two Doppler components. The MB-FTMW spectrum was obtained by averaging 200 experimental cycles (four free induction decay signals per cycle).

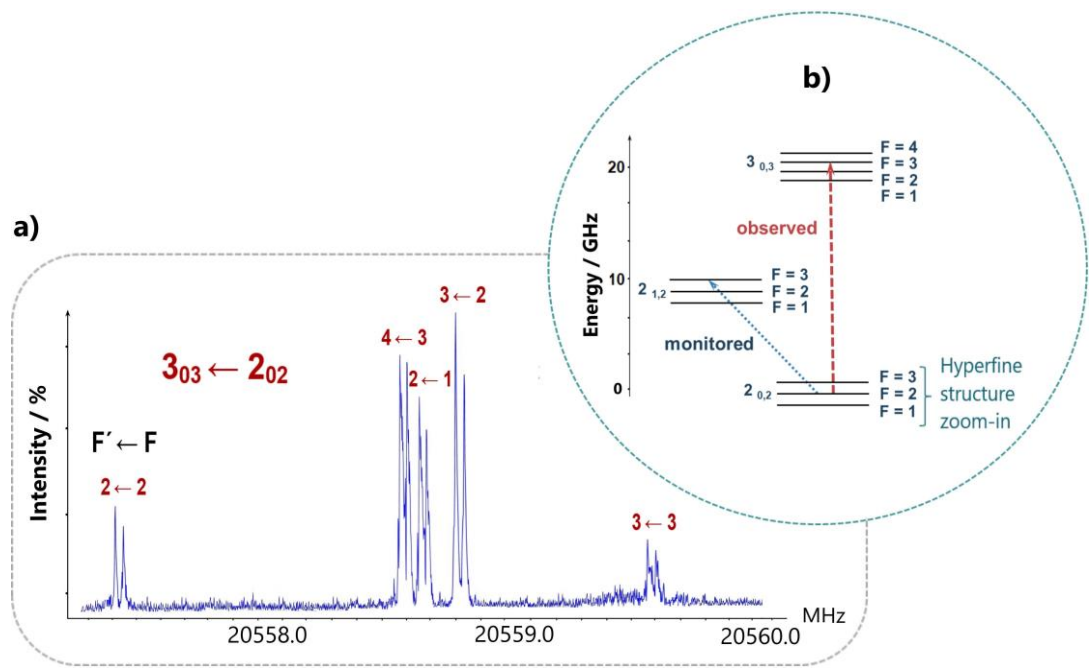

**Fig. 2.** (a) Double-resonance spectrum of the $3_{03} \leftarrow 2_{02}$ rotational transition of *syn*-glycolamide showing five different hyperfine components. The 2 2 and 3 2 lines were observed monitoring the hyperfine component $F = 2$ of the rotational transition $2_{12} \leftarrow 2_{02}$, while the 2 1 was observed monitoring the hyperfine component $F = 1$ component, and both 4 3 and 3 3 were observed monitoring the $F = 3$ component of the same rotational transition; (b) Schematic energy level diagram (in GHz) relative to the $2_{02}$ energy level of *syn*- glycolamide from the determined molecular constants. The hyperfine splittings are shown. Arrows indicate an example of the observed (dashed arrows in garnet) and monitored (dotted arrows in blue) transitions.

**Table 1.** Experimental spectroscopic parameters for *syn*- and *anti*- glycolamide.

| Parameters | *syn (DR)* | *anti* |
|---|---|---|
| A[a] (MHz) | 10454.2640 (15)[e] | 9852.6242 (22) |
| B (MHz) | 4041.15663 (54) | 4125.1664 (15) |
| C (MHz) | 2972.07394 (59) | 2967.57976 (75) |
| $\Delta_J$ (kHz) | 0.7723 (59) | 0.700 (38) |
| $\Delta_K$ (kHz) | 2.224 (74) | 3.45 (28) |
| $\Delta_{JK}$ (kHz) | 6.98 (46) | 3.3 (8) |
| $\delta_J$ (kHz) | 0.2061 (21) | 0.27 (2) |
| $\delta_K$ (kHz) | 2.082 (65) | - |
| $\chi_{aa}$[b] (MHz) | 2.050 (4) | 1.618 (5) |
| $\chi_{bb}$ (MHz) | 1.930 (6) | 2.020 (10) |
| $\chi_{cc}$ (MHz) | -3.981 (6) | -3.638 (10) |
| N[c] | 97 | 46 |
| $\sigma$[d] (kHz) | 21.3 | 15.6 |

**Notes.**[a] A, B, and C represent the rotational constants.[b] $\chi_{aa}$, $\chi_{bb}$, $\chi_{cc}$ are the diagonal elements of the $^{14}$N nuclear quadrupole coupling tensor.[c] N is the number of measured hyperfine components.[d] $\sigma$ is the root mean square (rms) deviation of the fit.[e] Standard error in parentheses in units of the last digit.

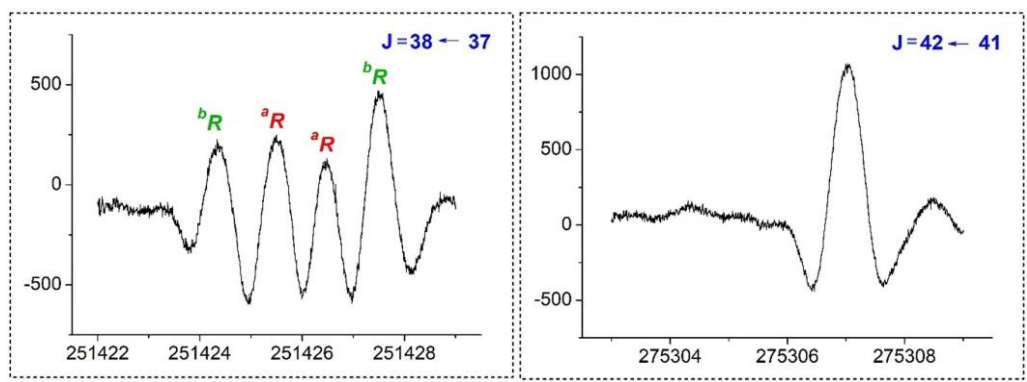

**Fig. 3.** Pairs of *a*- and *b*-type transitions of *syn*-glycolamide with $K_a = 4,5$ at *J''*=37 (left panel) that coa- lesce into a quadruply degenerate line at *J''*=41 (right panel) during the mmW experiment while going up in frequency (*b*-type transitions colored in green and *a*-type transitions in red, respectively). The broadening is higher than that of conventional room-temperature experiments due to the effect of the temperature of the cell. The *x* axis is labeled in MHz.

### 3.2. Millimeter and submillimeter wave spectra

Afterwards, predictions based on the spectroscopic constants from the high-resolution data of the previous section were used. Groups of *a*-type *R*-branch transitions dominate the millimeter-wave spectrum and can be easily identified due to their typical (*B+C*) periodicity. These groups are inter- spersed by *b*-type *R*-branch transitions that do not cluster into any characteristic pattern. However, as *J* increases, the rotational energy levels with the lowest $K_a$ quantum numbers become nearly degenerate, and pairs of *b*-type with corresponding *a*-type transitions involving these energy lev- els form one quadruply degenerate line. For instance, in Fig. 3 we depict the coalescence for the quartets corresponding to $K_a = 4$ and 5. Taking this into account, intense *aR*-branch transitions in- volving $K_a = 0$, 1 energy levels were searched for at first. Subsequently, higher $K_a$ R-branch *a*- and *b*- type transitions together with several *Q*-branch transitions were assigned. However, it is worth noticing that discrepancies up to 60 MHz were found at first between the experimental *Q*-branch and also the high $K_a$ rotational transitions compared to initial predictions, emphasizing once more the primary interest of this study.

Finally, 1480 new transition lines were measured (together with the previous MW and DR-MW measurements correctly weighted) and the ranges of *J* and $K_a$ quantum numbers were extended up to 51 and 33, respectively. The fits and predictions were made in terms of Watson's *A*-reduced Hamiltonian in $I^r$ -representation with the Pickett's SPFIT/SPCAT program suite (Pickett 1991). The presently available data allowed the determination of accurate rotational constants, the full set of quartic and almost all the sextic centrifugal distortion constants. What is more, despite the broadening due to the temperature effect of the cell, the RMS deviation of the fit is considerably small (around 50 kHz). The spectroscopic constants are presented in Table 2, and the measured transitions are reported in Table 4. The final spectroscopic set decisively improves the previous free-jet Stark modulated experimental data by up to three orders of magnitude (Maris 2004). In conclusion, the millimeter and submillimeter-wave spectrum of the most stable conformer of gly-

**Table 2.** Ground-state spectroscopic parameters of **syn**-glycolamide (A-Reduction, I$^r$-Representation).

| Parameters | This study | Stark experiment[a] |
|---|---|---|
| A[b] (MHz) | 10454.26296 (48)[f] | 10454.260 (6) |
| B (MHz) | 4041.15605 (24)[f] | 4041.163 (4) |
| C (MHz) | 2972.07260 (20) | 2972.083 (5) |
| $\Delta_J$ (kHz) | 0.77546 (14) | 0.80 (2) |
| $\Delta_K$ (kHz) | 2.35393 (55) | 2.36 (6) |
| $\Delta_{JK}$ (kHz) | 6.47368 (90) | 6.3 (2) |
| $\delta_J$ (kHz) | 0.204628 (23) | 0.202 (5) |
| $\delta_K$ (kHz) | 2.0742 (21) | 1.8 (2) |
| $\Phi_J$ (mHz) | -0.492 (36) | - |
| $\Phi_{JK}$ (Hz) | -0.02302 (28) | - |
| $\Phi_K$ (Hz) | 0.04078 (77) | - |
| $\varphi_{JK}$ (Hz) | -0.01456 (71) | - |
| $\varphi_K$ (Hz) | -0.1638 (21) | - |
| $\chi_{aa}$[c] (MHz) | 2.051 (4) | 2.3 (2) |
| $\chi_{bb}$ (MHz) | 1.930 (6) | 2.2 (1) |
| $\chi_{cc}$ (MHz) | -3.982 (6) | -3.5 (1) |
| N[d] | 1480 | 121 |
| $\sigma$[e] (kHz) | 50 | 34 |

**Notes.** [a] The parameters for the previous Stark experiment were published in Maris (2004). [b] A, B, and C represent the rotational constants. [c] $\chi_{aa}$, $\chi_{bb}$, $\chi_{cc}$ are the diagonal elements of the $^{14}$N nuclear quadrupole coupling tensor. [d] N is the number of measured transitions. [e] $\sigma$ is the root mean square (rms) deviation of the fit. [f] Standard error in parentheses in units of the last digit.

**Table 3.** Rotational and vibrational partition functions of glycolamide.

| Temperature (K) | $Q_r$[a] | $Q_v$[b] |
|---|---|---|
| 2.72 | 68.8745 | 1.0000 |
| 8.00 | 342.4694 | 1.0000 |
| 9.38 | 434.0770 | 1.0000 |
| 18.75 | 1224.6939 | 1.0024 |
| 37.50 | 3459.8737 | 1.0529 |
| 75.00 | 9781.5471 | 1.3302 |
| 100.00 | 15059.0883 | 1.6225 |
| 150.00 | 27667.5424 | 2.5317 |
| 225.00 | 50841.4225 | 5.2978 |
| 300.00 | 78299.5551 | 11.5094 |

**Notes.** [a] $Q_r$ is the rotational partition function. It does not take the hyperfine splitting into account. [b] $Q_v$ is the vibrational partition function. The total partition function of the molecule (without hyperfine splitting) is $Q_r \times Q_v$.

colamide has been studied up to 460 GHz to provide accurate laboratory reference spectra that are needed to compare directly against the observational data in different regions of the ISM.

We provide the rotational ($Q_r$) and vibrational ($Q_v$) partition functions of glycolamide in Table 3. The values of $Q_r$ were calculated from first principles at different temperatures, using the Picket program Pickett (1991). The maximum value of the $J$ quantum number of the energy levels taken into account to calculate the partition function is 200. The vibrational part, $Q_v$, was estimated using a harmonic approximation and a simple formula that corresponds to Eq. 3.60 of Gordy & Cook (1970). The frequencies of the normal modes were obtained from double-hybrid calculations of the harmonic force field using the B2PLYPD3 method and the aug-cc-pVTZ basis set (see Table A.2 of the Apendix). The values of $Q_v$ were calculated by taking the ten lowest vibrational modes into account. The full partition function, $Q_{tot}$, is thus the product of $Q_r$ and $Q_v$.

**Table 4.** Sample list of the measured transition frequencies for the ground-state of glycolamide.

| Technique | J' | $K_a'$ | $K_c'$ | F' | J'' | $K_a''$ | $K_c''$ | F'' | $\upsilon_{obs}{}^a$ (MHz) | $\upsilon_{calc}{}^b$ (MHz) | $\upsilon_{obs}$-$\upsilon_{calc}{}^c$ (MHz) |
|---|---|---|---|---|---|---|---|---|---|---|---|
| | 1 | 0 | 1 | 1 | 1 | 0 | 0 | 0 | 1 | 7013.7350 | 7013.7398 | 0.0048 |
| | 1 | 0 | 1 | 2 | 0 | 0 | 0 | 1 | 7013.1250 | 7013.1240 | 0.0009 |
| | 1 | 0 | 1 | 0 | 0 | 0 | 0 | 1 | 7012.2000 | 7012.2005 | 0.0005 |
| MB-FTMW | 2 | 0 | 2 | 1 | 1 | 0 | 1 | 1 | 13902.4690 | 13902.4652 | 0.0038 |
| | 2 | 0 | 2 | 3 | 1 | 0 | 1 | 2 | 13903.5250 | 13903.5242 | 0.0007 |
| | 2 | 0 | 2 | 1 | 1 | 0 | 1 | 0 | 13904.0050 | 13904.0038 | 0.0012 |
| | 2 | 0 | 2 | 2 | 1 | 0 | 1 | 1 | 13903.7090 | 13903.7073 | 0.0002 |
| | 2 | 0 | 2 | 2 | 1 | 0 | 1 | 2 | 13904.3250 | 13904.3228 | 0.0022 |
| | 3 | 0 | 3 | 2 | 2 | 0 | 2 | 2 | 20557.4200 | 20557.4348 | -0.0148 |
| | 3 | 0 | 3 | 4 | 2 | 0 | 2 | 3 | 20558.6000 | 20558.5948 | 0.0052 |
| DR-MB-FTMW | 3 | 0 | 3 | 2 | 2 | 0 | 2 | 1 | 20558.6900 | 20558.6769 | 0.0131 |
| | 3 | 0 | 3 | 3 | 2 | 0 | 2 | 2 | 20558.8291 | 20558.8291 | -0.0091 |
| | 3 | 0 | 3 | 3 | 2 | 0 | 2 | 3 | 20559.6400 | 20559.6276 | 0.0123 |
| | 17 | 0 | 17 | - | 16 | 1 | 16 | - | 103873.6006 | 103873.5786 | 0.0220 |
| | 17 | 1 | 17 | - | 16 | 1 | 16 | - | 103874.6090 | 103874.6102 | -0.0013 |
| | 17 | 0 | 17 | - | 16 | 0 | 16 | - | 103875.6817 | 103875.6598 | 0.0218 |
| | 17 | 1 | 17 | - | 16 | 0 | 16 | - | 103876.7134 | 103876.6915 | 0.0219 |
| | 38 | 5 | 34 | - | 37 | 4 | 33 | - | 251427.5135 | 251427.5373 | -0.0239 |
| | 38 | 4 | 34 | - | 37 | 4 | 33 | - | 251426.4797 | 251426.4907 | -0.0109 |
| mmW | 38 | 5 | 34 | - | 37 | 5 | 33 | - | 251425.4969 | 251425.4909 | 0.0060 |
| | 38 | 4 | 34 | - | 37 | 5 | 33 | - | 251424.3507 | 251424.3287 | 0.0219 |
| | 48 | 31 | 17 | - | 48 | 30 | 18 | - | 419757.5597 | 419757.5203 | 0.0394 |
| | 48 | 31 | 18 | - | 48 | 30 | 19 | - | 419757.5597 | 419757.5203 | 0.0394 |
| | 49 | 31 | 18 | - | 49 | 30 | 19 | - | 419568.2608 | 419568.2301 | 0.0307 |
| | 49 | 31 | 19 | - | 49 | 30 | 20 | - | 419568.2608 | 419568.2301 | 0.0307 |
| | 32 | 17 | 15 | - | 31 | 16 | 16 | - | 453552.8166 | 453552.8010 | 0.0155 |
| | 32 | 17 | 16 | - | 31 | 16 | 15 | - | 453552.8166 | 453552.8010 | 0.0156 |

**Notes.** Upper and lower state quantum numbers are indicated by ' and '', respectively.[a] Observed frequency.[b] Calculated frequency.[c] Observed minus calculated frequency. Table A3 containing the com- plete list of measured transitions is only available in electronic form at the CDS via anonymous ftp to cdsarc.u-strasbg.fr (130.79.128.5) or via http://cdsweb.u-strasbg.fr/cgi-bin/qcat?J/A+A/.

# 4. Search for glycolamide toward Sgr B2(N2) with ALMA

## 4.1. Observations

We used the spectroscopic predictions derived in Sect. 3 to search for glycolamide toward the hot molecular core Sgr B2(N2) located in the giant molecular cloud Sgr B2. We used the EMoCA imaging spectral line survey performed toward the protocluster Sgr B2(N) with ALMA. As mentioned in Sect. 1, Sgr B2(N) is a high-mass star forming region with a very rich chemistry that is part of the Sagittarius B2 Giant Molecular Cloud in which many complex organic molecules were discovered for the first time in the interstellar medium (see, e.g., Menten 2004, and Sect. 5.3). A detailed account of the observational setup and data reduction of the EMoCA survey can be found in Belloche et al. (2016). In short, the survey covers the frequency range from 84.1 GHz to 114.4 GHz with a spectral resolution of 488 kHz (1.7 to 1.3 km s$^{-1}$) and a median angular resolution of 1.6″. The field was centered at $(\alpha, \delta)_{J2000}$= $(17^h47^m19\overset{s}{.}87, -28°22'16\overset{''}{.}0)$. Here we analyze the spectrum at the peak position of Sgr B2(N2) located at $(\alpha, \delta)_{J2000}$= $(17^h47^m19\overset{s}{.}86, -28°22'13\overset{''}{.}4)$.

As described in Belloche et al. (2016), the spectrum was modeled with the software Weeds (Maret et al. 2011) under the assumption of local thermodynamical equilibrium (LTE), which is an appropriate assumption given the high densities characterizing the region where the hot-core emission is detected ($> 1 \times 10^7 \, \mathrm{cm}^{-3}$, see Bonfand et al. 2019). A best-fit synthetic spectrum of each molecule was derived separately, and the contributions of all identified molecules were then added together. Each species was modeled with a set of five parameters: size of the emitting region, column density, temperature, linewidth, and velocity offset with respect to the assumed systemic velocity of the source ($74 \, \mathrm{km \, s^{-1}}$).

### 4.2. Nondetection of glycolamide

We did not find a clear evidence for the presence of glycolamide in Sgr B2(N2). Figure 4 shows in red the synthetic spectrum that we computed to obtain the upper limit on the glycolamide column density. In order to model the emission of glycolamide, we assumed the same parameters as those derived by Belloche et al. (2017) for acetamide ($CH_3C(O)NH_2$): an emission size of $0.9''$, a temperature of 180 K, a linewidth of $5 \, \mathrm{km \, s^{-1}}$, and no velocity offset with respect to the systemic velocity. The red spectrum shown in Fig. 4 was computed with a glycolamide column density of $2.4 \times 10^{16} \, \mathrm{cm}^{-2}$ (see Table 5).

**Table 5.** Parameters of our best-fit LTE model of acetaldehyde, glycolaldehyde, and acetamide, and column density upper limit for glycolamide, toward Sgr B2(N2).

| Molecule | Status[a] | $N_{det}$[b] | Size[c] ($''$) | $T_{rot}$[d] (K) | $N$[e] ($\mathrm{cm}^{-2}$) | $F_{vib}$[f] | $\Delta V$[g] ($\mathrm{km \, s^{-1}}$) | $V_{off}$[h] ($\mathrm{km \, s^{-1}}$) | $\frac{N_{ref}}{N}$[i] |
|---|---|---|---|---|---|---|---|---|---|
| **Acetaldehyde** | | | | | | | | | |
| $CH_3CHO$, $r = 0$[γ] | d | 19 | 1.1 | 160 | 5.3 (17) | 1.08 | 5.6 | 0.0 | 1 |
| $\quad r_t = 1$ | d | 13 | 1.1 | 160 | 5.3 (17) | 1.08 | 5.6 | 0.0 | 1 |
| $\quad r_t = 2$ | d | 2 | 1.1 | 160 | 5.3 (17) | 1.08 | 5.6 | 0.0 | 1 |
| $^{13}CH_3CHO$, $r = 0$ | t | 2 | 1.1 | 160 | 2.4 (16) | 1.01 | 5.6 | 0.0 | 22 |
| $\quad r_t = 1$ | n | 0 | 1.1 | 160 | 2.4 (16) | 1.01 | 5.6 | 0.0 | 22 |
| $CH_3^{13}CHO$, $r = 0$ | n | 0 | 1.1 | 160 | 2.4 (16) | 1.01 | 5.6 | 0.0 | 22 |
| $\quad r_t = 1$ | n | 0 | 1.1 | 160 | 2.4 (16) | 1.01 | 5.6 | 0.0 | 22 |
| **Glycolaldehyde** | | | | | | | | | |
| $CH_2(OH)CHO$, $r = 0$[γ] | d | 21 | 1.0 | 170 | 1.1 (17) | 0.92 | 5.4 | 1.0 | 1 |
| $\quad r_1 = 1$ | t | 2 | 1.0 | 170 | 1.1 (17) | 0.92 | 5.4 | 1.0 | 1 |
| $\quad r_2 = 1$ | t | 1 | 1.0 | 170 | 1.1 (17) | 0.92 | 5.4 | 1.0 | 1 |
| $^{13}CH_2(OH)CHO$, $r = 0$ | n | 0 | 1.0 | 170 | < 1.6 (16) | 1.37 | 5.4 | 1.0 | > 6.7 |
| $CH_2(OH)^{13}CHO$, $r = 0$ | n | 0 | 1.0 | 170 | < 2.1 (16) | 1.37 | 5.4 | 1.0 | > 5.4 |
| **Acetamide** | | | | | | | | | |
| $CH_3C(O)NH_2$, $r = 0$[γ] | d | 10 | 0.9 | 180 | 1.4 (17) | 1.23 | 5.0 | 1.5 | 1 |
| $\quad r_t = 1$ | d | 8 | 0.9 | 180 | 1.4 (17) | 1.23 | 5.0 | 1.5 | 1 |
| $\quad r_t = 2$ | d | 5 | 0.9 | 180 | 1.4 (17) | 1.23 | 5.0 | 1.5 | 1 |
| $\quad \Delta r_t \ç\ 0$ | t | 0 | 0.9 | 180 | 1.4 (17) | 1.23 | 5.0 | 1.5 | 1 |
| **Glycolamide** | | | | | | | | | |
| $CH_2(OH)C(O)NH_2$, $r = 0$ | n | 0 | 0.9 | 180 | < 2.4 (16) | 3.38 | 5.0 | 0.0 | – |

**Notes.** The parameters for acetamide were published in Belloche et al. (2017). [a] d: detection, t: tentative detection, n: non-detection. [b] Number of detected lines (conservative estimate, see Sect. 3 of Belloche et al. 2016). One line of a given species may mean a group of transitions of that species that are blended together. [c] Source diameter ($FWHM$). [d] Rotational temperature. [e] Total column density of the molecule. $x$ ($y$) means $x \times 10^y$. An identical value for all listed vibrational and torsional states of a molecule means that LTE is an adequate description of the vibrational and torsional excitation. [f] Correction factor that was applied to the column density to account for the contribution of vibrationally excited states, in the cases where this contribution was not included in the partition function of the spectroscopic predictions. For glycolaldehyde, see explanation in Sect. 4.3. [g] Linewidth ($FWHM$). [h] Velocity offset with respect to the assumed systemic velocity of Sgr B2(N2), $V_{sys} = 74 \, \mathrm{km \, s^{-1}}$. [i] Column density ratio, with $N_{ref}$ the column density of the previous reference species marked with a γ.

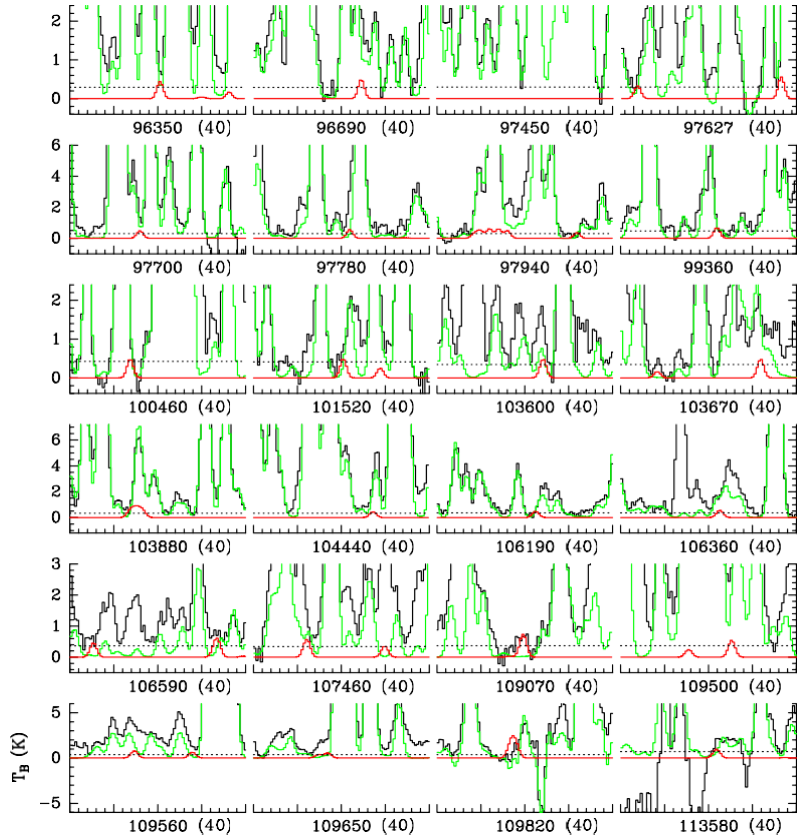

**Fig. 4.** Synthetic LTE spectrum of glycolamide (in red) used to derive the upper limit to its column density, overlaid on the ALMA spectrum of Sgr B2(N2) (in black) and the synthetic spectrum that contains the contributions of all the species (but not glycolamide) that we have identified so far in this source (in green). The dotted line in each panel indicates the $3\sigma$ noise level. For clarity purposes, the $x$-axis of each panel is only labeled with its central frequency followed by the value of the displayed frequency range in parentheses, both in MHz. The transitions of glycolamide that are weaker than $3\sigma$ and the ones which are heavily contaminated by other species are not shown.

The ALMA spectrum displayed in Fig. 4 shows a $\sim$5$\sigma$ line matching one line of the synthetic spectrum of glycolamide at $\sim$109070 MHz, but several issues at other frequencies suggest that this is likely fortuitous. One major problem is indeed that the glycolamide synthetic spectrum at $\sim$103875 MHz, if added to the contribution of all identified species, would significantly overestimate the signal detected with ALMA at this frequency, which is already accounted for (even slightly overestimated) by the contributions of ethanol and $CH_2^{13}CHCN$. The latter species both have many lines detected in the survey and well reproduced by our best-fit LTE model (see Belloche et al. 2016; Müller et al. 2016). There is thus no room left for glycolamide emission at this frequency.

Another problem occurs at $\sim$109817 MHz where the red spectrum overestimates the detected signal. However, as can be seen in the green spectrum, there are several absorption components tracing diffuse and translucent clouds along the line of sight that are expected from $C^{18}O$ $J$ =1–0 in this frequency range (see, e.g., Thiel et al. 2019), in particular two of them very close to the glycolamide transition frequency. The strength of these $C^{18}O$ absorption components is difficult to constrain. Therefore, the inconsistency of the glycolamide spectrum at this frequency may not be significant.

All in all, the fact that only one line of glycolamide matches a detected line and the fact that there is no room left for glycolamide emission at ~103875 MHz lead us to conclude that glycolamide is not detected in Sgr B2(N2). We thus consider the red spectrum shown in Fig. 4 as an upper limit to the emission of glycolamide in Sgr B2(N2).

### 4.3. Comparison to other molecules

In order to put the column density upper limit derived for glycolamide into a broader astrochemical context, we use the EMoCA survey to determine the column densities of other species detected toward Sgr B2(N2) that can be compared to glycolamide. We selected acetaldehyde ($CH_3CHO$), glycolaldehyde ($CH_2(OH)CHO$), and acetamide ($CH_3C(O)NH_2$) because acetamide and glycolamide are structurally related in the same way as acetaldehyde and glycolaldehyde: in both cases, the substitution of an hydrogen atom in the methyl group of the former species with an hydroxyl group gives the latter species. The column density derived for acetamide in Sgr B2(N2) was already reported in Belloche et al. (2017).

For both acetaldehyde and glycolaldehyde, we use the spectroscopic predictions (entries 44003 version 3 and 60006 version 2, respectively) available in the Jet Propulsion Laboratory spectroscopic catalog (Pickett et al. 1998). The acetaldehyde entry is based on Kleiner et al. (1996) with data in the range of our survey from Kleiner et al. (1991, 1992). The ground state data of glycolaldehyde are based on Butler et al. (2001), those of the excited data on Widicus Weaver et al. (2005). Recently, we became aware of an extensive study of the vibrational spectrum of glycolaldehyde by Johnson et al. (2013). The vibrational energies are higher than those used for the JPL catalog which were based on relative intensities of vibrational states in the rotational spectrum of glycolaldehyde. As a result, the vibrational correction factor to the partition function at 170 K is only 1.38 instead of 1.50. We made the appropriate correction to the column density derived below. For the $^{13}C$ isotoplogs of acetaldehyde, we use spectroscopic predictions from Margulès et al. (2015). For the $^{13}C$ isotopologs of glycolaldehyde, we use predictions (entries 61513 and 61514, both version 1) available in the Cologne Database for Molecular Spectroscopy (Endres et al. 2016). The $^{13}C$ entries of glycolaldehyde are based on Haykal et al. (2013). The acetamide entry is based on Ilyushin et al. (2004).

Many transitions of acetaldehyde and glycolaldehyde in their vibrational ground state are detected toward Sgr B2(N2) (see Figs. B.1 and B.8). We also detect transitions from within the first two torsionally excited states of acetaldehyde (Figs. B.2 and B.3) and transitions from within the first two vibrationally excited states of glycolaldehyde (Figs. B.9 and B.10). There is an issue with the transition of $CH_3CHO$ $r_t = 2$ at ~90910 MHz, and we suspect that this results from an inaccurate rest frequency. The frequency computed for this transition by Kleiner et al. (1996) is indeed somewhat higher than the frequency listed in the JPL catalog: $90910.914 \pm 0.204$ MHz versus $90910.330 \pm 0.101$ MHz. The former would better match the EMoCA spectrum.

**Table 6.** Rotational temperatures derived from population diagrams toward Sgr B2(N2).

| Molecule | States[a] | $T_{\mathrm{fit}}$[b] (K) |
|---|---|---|
| CH$_3$CHO | $r = 0$, $r_{\mathrm{t}} = 1$, $r_{\mathrm{t}} = 2$ | 166 (17) |
| CH$_2$(OH)CHO | $r = 0$, $r_1 = 1$, $r_2 = 1$ | 163.8 (6.1) |

**Notes.** [a] Vibrational states that were taken into account to fit the population diagram. [b] The standard deviation of the fit is given in parentheses. As explained in Sect. 3 of Belloche et al. (2016), this uncertainty is purely statistical and should be viewed with caution. It may be underestimated.

On the basis of a first-guess LTE synthetic spectrum for each molecule, we selected the transitions that are not significantly contaminated by other species to measure the size of the emission of acetaldehyde and glycolaldehyde. The emission is compact around Sgr B2(N2) and we obtained median sizes (HPBW) of 1.1$^{IJ}$ and 1.0$^{IJ}$, respectively. The transitions that are not much contaminated were then used to produce population diagrams (Figs. C.1a and C.2a). Following the method described in Belloche et al. (2016), we corrected the diagrams for optical depth and for the contamination due to the other species using our synthetic spectrum that includes the contribution of all species identified so far (Figs. C.1b and C.2b). A fit to the corrected population diagrams yields temperatures of 166 ± 17 K and 164 ± 6 K for acetaldehyde and glycolaldehyde, respectively (see Table 6). As explained in Sect. 3 of Belloche et al. (2016), the uncertainties on the rotational temperatures are purely statistical and do not take into account the residual contamination by species that have not been identified so far. The best-fit spectra shown in Figs. B.1–B.3 and Figs. B.8–B.10 were computed with temperatures of 160 K and 170 K, respectively, consistent within the uncertainties with the values derived from the fits to the population diagrams.

On the basis of the parameters obtained for acetaldehyde and glycolaldehyde, we also searched for emission of their $^{13}$C isotopologs, assuming a $^{12}$C/$^{13}$C isotopic ratio in the range 20–25 as usually found for complex organic molecules in Sgr B2(N2) (see, e.g., Belloche et al. 2016; Müller et al. 2016). We tentatively detect $^{13}$CH$_3$CHO in its vibrational ground state (Fig. B.4), with an isotopic ratio of 22. Assuming the same isotopic ratio for both $^{13}$C isotopologs, which is commonly the case in Sgr B2(N2) for complex organic molecules (COMs) that contain several carbon atoms (Belloche et al. 2016; Margulès et al. 2016; Müller et al. 2016), yields a synthetic spectrum of CH$_3$$^{13}$CHO that is consistent with the observed spectrum (Fig. B.6). However, no transition is sufficiently free of contamination from emission of other species to claim a detection of this isotopolog. Still, we include it in our complete model because its emission contributes significantly to the signal detected at several frequencies. Transitions from the first torsionally excited state of both isotopologs contribute to the observed spectra at the ~3σ level at most (see Figs. B.5 and B.7). The $^{13}$C isotopologs of glycolaldehyde are not detected. We report upper limits to their column densities in Table 5. Their non-detection is consistent with the $^{12}$C/$^{13}$C typical ratio obtained for other COMs in Sgr B2(N2).

From Table 5 we derive the following column density ratios in Sgr B2(N2): [CH$_3$CHO]/[CH$_2$(OH)CHO] = 4.8 and [CH$_3$C(O)NH$_2$]/[CH$_2$(OH)C(O)NH$_2$] > 5.8. The lower limit on the latter ratio is only slightly higher than the former ratio. We also find that glycolamide ist at least five times less abundant than glycolaldehyde ([CH$_2$(OH)CHO]/[CH$_2$(OH)C(O)NH$_2$] > 4.6).

## 5. Search for glycolamide toward Sgr B2(N) with Effelsberg

At low radio frequencies, in the cm-wavelength range, transitions of many molecules including complex organic species are often seen in absorption against the strong nonthermal (synchrotron) radiation (Hollis et al. 2007) of the Sgr B2 comples and also the greater Galactic center region, while others appear in emission depending on the excitation process (e.g., Menten 2004; Hollis et al. 2004b, 2006, and see Sect. 5.3 below) . In the following we describe our search for several low-$J$ transitions of glycolamide within in the 4–8 and 12–17 GHz frequency ranges.

### 5.1. Observations

We performed measurements (project id: 08-20) toward Sgr B2(N) with the S45mm and S20mm receivers of the 100 m radio telescope at Effelsberg, Germany[1] on 28 March and 3 April 2020. The targeted position was $\alpha_{J2000}$=17$^h$47$^m$19.8$^s$, $\delta_{J2000}$=−28°22'17''. This is identical to the position used by Hollis et al. (2006) for their acetamide observations and within a few arc seconds of the position of Sgr B2(N2) hot core for which the ALMA spectra discussed in Sect. 4 were extracted. We used the position-switching mode with an off-source position located 60' east of Sgr B2(N). The S45 mm receiver is a broad-band receiver with two orthogonal linear polarizations operating in the frequency range between 4 and 9.3 GHz. The S20 mm receiver is a dual-beam receiver with two orthogonal linear polarizations covering the frequency range 12–18 GHz.

The focus was adjusted using observations of 3C286 at the beginning of each observing session. 3C286 was also used as flux calibrator, and the flux calibration accuracy is estimated to be within ∼10%. NRAO530 was used as pointing source close to Sgr B2(N). The pointing accuracy was found to be better than 5'', small compared to the half-power beam width (HPBW) of 99''at 7.2 GHz and 57''at 13 GHz. The main beam efficiency is about 69% at 7.2 GHz and 66% at 13 GHz, and the typical system temperature was about 33 K at 7.2 GHz and 41 K at 13 GHz.

The spectroscopic backends were Fast Fourier Transform spectrometers (FFTSs, e.g., Klein et al. 2012). The FFTSs cover, in two bands, the frequency ranges 4–8 GHz and 12–17 GHz with a channel width of 38.1 kHz, corresponding to a velocity spacing of 1.6 km s$^{-1}$ at 7 GHz and 0.9 km s$^{-1}$ at 13 GHz. This is adequate, given the expected line widths of ∼10 km s$^{-1}$ (see below). The data reduction was performed using the GILDAS/CLASS software[2].

### 5.2. Nondetection of glycolamide

We used the spectroscopic predictions (with hyperfine structure) derived in Sect. 3 to search for glycolamide in the Effelsberg spectra of Sgr B2(N). None of the glycolamide transitions covered by our observations were detected either in absorption or in emission. Figures 5 and B.11 illustrate these non-detections with several portions of the spectra that cover some of the lowest-energy transitions of glycolamide included in our observations. The spectra of Figs. 5a-b and B.11a-d were obtained after averaging both polarizations. We then measured the average continuum level in each panel and removed a fifth-order baseline to produce the spectra shown in Figs. 5c-d and B.11e-h.

We used the software Weeds to produce synthetic spectra assuming a beam-filling factor of 1. We made the same assumptions as Hollis et al. (2006) made to analyze their detection of acetamide with the GBT: a systemic velocity of 64 km s$^{-1}$ and a rotational temperature of 8 K. We assumed a linewidth (FWHM) of 10 km s$^{-1}$, which is similar to the linewidth of glycolaldehyde detected with the GBT (Hollis et al. 2004b). We assumed a background temperature equal to the average continuum level determined in Figs. 5a-b and B.11a-d. With these parameters, we obtained a column density upper limit of $1.2 \times 10^{14}$ cm$^{-2}$ for glycolamide at the 4–5$\sigma$ level for extended emission from this molecule. This value is similar to the column densities derived by Hollis et al. (2004b, 2006) for acetamide and glycolaldehyde with the GBT.

For completeness, Table A.0 lists all glycolamide transitions with a lower level energy in temperature units lower than 10 K covered by the Effelsberg observations. None of these transitions are detected, and the table indicates the noise level and the average continuum level.

### 5.3. Discussion of the Effelsberg nondetection

Our astronomical search for centimeter wavelength transitions from glycolamide was motivated by the fact that in the early days of molecular radio astronomy, the early 1970s, radiation from many large molecules was first detected toward Sgr B2 with single dish telescopes (for a sum- mary and a list of early detections, see Menten 2004). These identifications were made via de- tections of cm-wavelength of low-$J$ rotational ground-state or near ground state transitions. More recent observations with the GBT added a significant number of species to this list, namely glyco-laldehyde (CH$_2$(OH)CHO), ethylene glycol (HOCH$_2$CH$_2$OH), propenal (CH$_2$CHCHO), propanal (CH$_3$CH$_2$CHO) and acetamide (CH$_3$C(O)NH$_2$) (Hollis et al. 2002, 2004a,b, 2006).

In many cases, these low-$J$ transitions were exclusively found toward Sgr B2 and its larger environment, the central molecular zone (CMZ) of our Galaxy (Morris & Serabyn 1996). The CMZ is a $\sim 200$ pc long region stretched around the Galactic center that contains molecular gas that has much higher temperatures and densities than Giant Molecular Clouds (GMCs) in the Galactic disk as well as a peculiar chemistry. The latter two facts cause a variety of molecules to be observable over the whole CMZ, via mm wavelength rotational lines (e.g., Jones et al. 2012; Requena-Torres et al. 2006), but also in the 834.3 MHz $1_1 - 1_1 A^{\mp}$ line of CH$_3$OH and low-J lines from various COMs (Gottlieb et al. 1979; Requena-Torres et al. 2008).

The star formation/GMC complex Sagittarius B2 is a prominent part of the CMZ and the men- tioned low-$J$ lines were first (and many of them only) detected toward the extended molecular envelope of this source, which is a prominent source of mm-wavelength emission from very many molecules (e.g., Jones et al. 2007). Important here is the imaging of the 1065 MHz $1_{11} - 1_{10}$ rotational transition of acetaldehyde ($CH_3CHO$), which shows its emission to be extended over the whole complex, is, a $3 \times 10$ pc$^2$ sized region (Chengalur & Kanekar 2003) that appears to be part of the several times larger extended envelope of the Sgr B2 complex (Comito et al. 2010; Schmiedeke et al. 2016). The even (much) more extreme conditions in its embedded dense molecular cores, for example, Sgr B2(N2), allow the detection of many rare species in these very compact regions that have sizes of order 1" (or $\approx 0.04$ pc), for instance, with the EMoCA survey (see Table 5). For the column densities and sizes reported in Table 5, any thermal (LTE) signal in the cm lines for the listed species would be undetectable because of the lines' low transition probabilities and the large beams they are observed with. In fact, the Weeds LTE model of Sgr B2(N2) (Sect. 4) produces a peak temperature of a few µ$K$ for the 7013 MHz lines of glycolamide when computed at the resolution of the Effelsberg telescope.

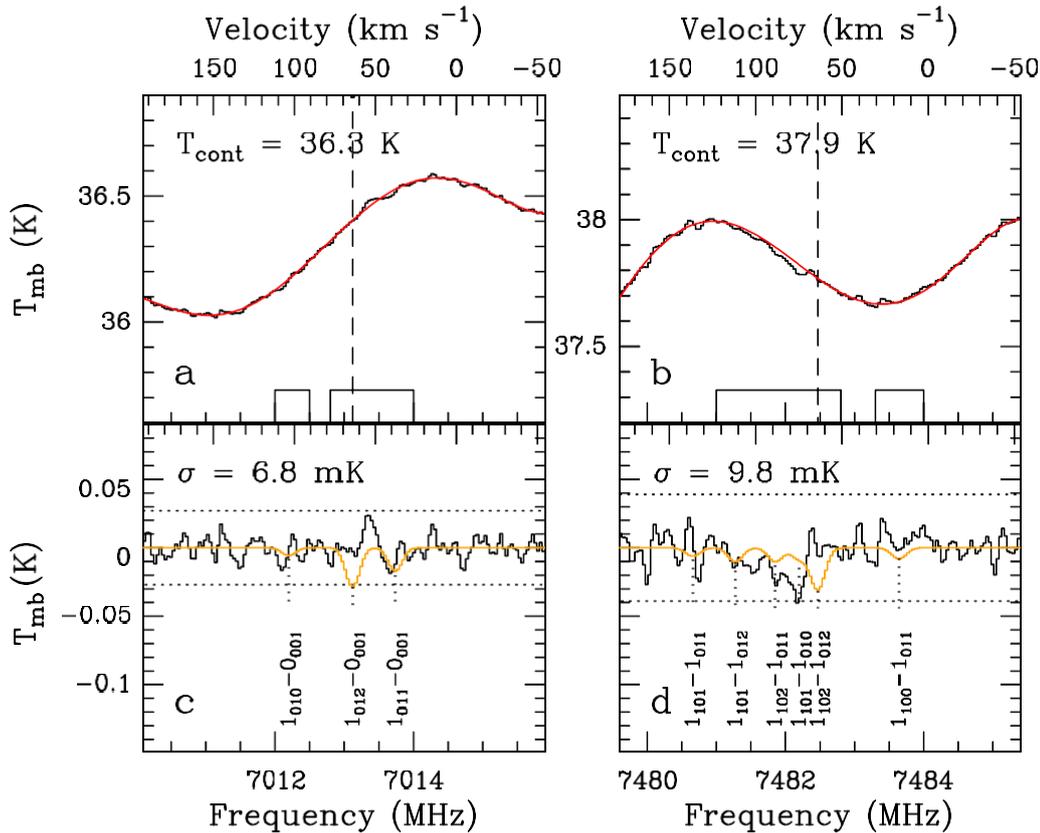

**Fig. 5. a, b** Examples of spectra obtained with Effelsberg between 4 and 8 GHz toward Sgr B2(N) resulting from averaging both polarizations. In each panel, the vertical dashed line marks a velocity of 64 km s$^{-1}$. The red curve shows the baseline fitted with a $5^{\text{th}}$-order polynomial function. The boxes at the bottom show the windows avoided to perform the baseline fit. The average continuum level used to compute the synthetic spectrum is indicated in the top left corner. **c, d** Synthetic spectra (orange) used to derive the upper limit to the column density of glycolamide, overlaid on observed spectra after baseline subtraction (black). The rms noise level is indicated in the top left corner. The horizontal dotted lines show the $4\sigma$ and $-4\sigma$ levels. The components of each hyperfine multiplet are labeled and marked with vertical dotted lines. The velocity axis shown at the top was computed relative to the strongest component of each hyperfine multiplet.

For various reasons, it is clear that the low frequency, low-$J$, emission or absorption lines observed from some molecules in the CMZ are only observable because their observed intensities are the result of a population inversion causing weak maser emission in some lines or *anti*-inversion causing enhanced absorption (over-cooling) in others. The weak (*anti*-)inversion results in generally small (absolute) optical depths. Nevertheless, the prominent background continuum emission causes the weak maser emission or absorption to be detectable, since the line signal is proportional to the optical depth times the (very strong) continuum flux. To give an example: that non-thermal excitation produces the observed intensities of the formamide and acetamide lines reported by Hollis et al. (2006) follows directly from the facts that, first, some of them show absorption, while others show emission. Second the total column densities calculated from the upper level column densities determined for individual lines vary by a factor of six for the acetamide lines and a factor of ten for the formamide lines, indicating non-LTE conditions. To calculate these column densities, Hollis et al. (2006) assume a rotation temperature of 8 K, which suggests a widespread "cold envelope" as the origin of these molecules, which is also implied by GBT observations of other COMs (see references in Hollis et al. 2006). Also Requena-Torres et al. (2006) find mm-wavelength emission from many COMs to be widespread over the CMZ and characterized by a rotation temperature of 8 K, while the same group of authors (Requena-Torres et al. 2008) derive 8–16 K from the cm lines of assorted COMs. They invoke subthermal excitation to explain these low numbers, which are much lower than the gas kinetic temperatures of >60 K found throughout the CMZ by Ginsburg et al. (2016), who find the extended Sgr B2 cloud to have some of the highest values (>100 K). For consistency with other studies of cm lines, We have adopted 8 K to calculate our column density upper limits.

Whether inversion or *anti*-inversion (over-cooling) of a line can be attained, depends on the interplay of the molecule's energy level structure and its radiative and collisional (de)excitation rates. Since for most species the latter are unknown, it is in general impossible to predict or explain observed maser (or anti-maser) action. Recently, collisional rate coefficients were calculated for methyl formate ($CH_3OCHO$) and methanimine ($CH_2NH$) by Faure et al. (2014) and Faure et al. (2018), respectively. Using their results, these authors' excitation calculations indeed confirm non-LTE excitation and reproduce the inversion observed in several low-J lines of these molecules.

While some low-J lines from the above-mentioned COMs and assorted others show extended non-thermal weak maser emission or enhanced absorption, others do not. Obviously, glycolamide does not and neither does amino acetonitrile, according to Brown et al. (1977) and Belloche et al. (2008), who unsuccessfully searched for this molecule's $1_{01} - 0_{00}$ multiplet at 9071.28 MHz (but did detect its mm-wavelength spectrum in the Sgr B2(N) hot core).

The upper limit we obtained from the cm-wavelength data in Sect. 5.2, $1.2 \times 10^{14} \, \text{cm}^{-2}$, is two orders of magnitude lower than the upper limit we calculated for the glycolamide column density in the Sgr B2(N2) hot core, $2.4 \times 10^{16} \, \text{cm}^{-2}$ (Table 5). These two values cannot be compared directly because the $H_2$ column densities of the hot core and Sgr B2(N)'s envelope probed with ALMA and Effelsberg, respectively, are widely different. We note that the value for Sgr B2(N2) is calculated from a well-constrained measurement, while the value for the envelope depends on the questionable assumption of LTE with a rotation temperature of 8 K. Would we assume a value of 100 K similar to the kinetic temperature of the envelope (Ginsburg et al. 2016), the upper limit based on the cm data would increase by a factor of 43 to $5.3 \times 10^{15}$ cm$^{-2}$. The acetamide column density presented by Hollis et al. (2006) would increase by a similar factor. Assuming 100 K for the rotation temperature is taking the opposite extreme to 8 K and is also an unrealistic assumption. The density in the molecular envelope is estimated to be of order a few times $10^3$ cm$^{-3}$ (Comito et al. 2010). At such moderate densities, the level population of any molecule with a substantial dipole moment will not be in LTE. Column density and abundance determinations for this environment are very difficult to make on the basis of a limited number of observed transitions, and impossible if their excitation is affected by inversion or *anti*-inversion.

## 6. Astrochemical modeling

In order to interpret the observational findings of Sect. 4, here we present results from the astro-chemical kinetics model *MAGICKAL* relating to the cold collapse and gradual warm up (8–400 K) of a hot core with physical conditions appropriate to Sgr B2(N2). Selected model results were originally published by Garrod et al. (2017), who extended their chemical network to simulate the production of butyl cyanide and related species. The results presented here for other molecules are produced by the medium-timescale warm-up model described in detail by those authors. In brief, the physical model involves a freefall collapse from $n_H = 3 \times 10^3$ to $2 \times 10^8$ cm$^{-3}$ over a period of $\sim 1$ Myr. The warm-up stage then proceeds over a period $2.85 \times 10^5$ yr, reaching a temperature of 200 K after $2 \times 10^5$ yr. The chemical model includes gas-phase, grain-surface, and ice-mantle chemistry. In the latter two phases in particular, the addition of functional-group radicals produced by cosmic ray-induced UV photolysis can result in a range of complex organic molecules, prior to the thermal desorption of the ice mantles into the gas phase.

Figure 6 shows the simulated time-dependent fractional abundances (with respect to H$_2$) of the same molecules whose observational parameters are shown in Table 5; both the gas-phase and grain-surface abundances are indicated for each species. Table 7 shows the peak gas-phase fractional abundances and the associated gas and dust temperatures obtained from the model for each of these species and some related molecules. In order to compare with the observational data, the ratios between the peak fractional abundances of key molecules are considered.

The ratio of glycolamide to glycolaldehyde found in the models (0.21) is consistent with the upper limit of the observational ratio (0.22). In the models, the main precursor to glycolamide is indeed glycolaldehyde; the latter is produced on the grains at around 30 K through the addition of the HCO and CH$_2$OH radicals, which are methanol photo-products. The abstraction of a hydrogen atom from the aldehyde end of the glycolaldehyde molecule, by various species including the radicals NH$_2$ and OH, produces a radical that may combine with NH$_2$ in the ice, or with NH followed by atomic hydrogen, resulting in glycolamide.

**Table 7.** Peak fractional abundances for selected molecules with respect to $H_2$ obtained from the chemical kinetics model. The associated temperature at which the peak value is achieved is also indicated.

| Molecule | n(i) / n($H_2$) | T (K) |
|---|---|---|
| $CH_3OH$ | 2.1(-5) | 130 |
| $CH_3CHO$ | 3.3(-7) | 58 |
| $CH_2(OH)CHO$ | 7.4(-7) | 148 |
| $CH_3C(O)NH_2$ | 3.4(-8) | 138 |
| $CH_2(OH)C(O)NH_2$ | 1.5(-7) | 226 |
| $NH_2CHO$ | 4.8(-7) | 158 |
| $CH_3OCHO$ | 7.8(-8) | 125 |

**Notes.** x (y) means $x \times 10^y$.

In the models, the main precursor to acetamide is its smaller homologue, formamide. Similarly to the grain-surface mechanism described above for glycolamide, hydrogen abstraction from the aldehyde end of formamide results in a radical that, with the addition of the methyl radical ($CH_3$), produces acetamide. The model ratio of acetamide to formamide (0.071) is comparable to the observational ratio (0.054), based on data from Table 5 and from Belloche et al. (2019, their Table 6).

While the ratios between each molecule and its precursor appear to agree fairly well with the observational values for glycolamide and acetamide, the direct ratio between these two species themselves does not; the models suggest a ratio of glycolamide to acetamide of 4.4, compared with an observational upper limit of 0.17, discrepant by a factor ∼26. The modeled ratios of the precursors of these molecules show similar behavior, being too large by a factor ∼33. This suggests that, while the chemistry of glycolamide and acetamide are well treated within the model, the relative abundances of their precursors require attention. It may be noted that the model abundance of methyl formate ($CH_3OCHO$), also shown in Table 7, is around one order of magnitude lower than that of glycolaldehyde. This also defies the observational expectation for Sgr B2(N2) and various other star-forming regions that typically show the inverse relationship; Belloche et al. (2016) found a column density of methyl formate of $1.2 \times 10^{18}$ cm$^{-2}$, ten times larger than the glycolaldehyde abundance shown in Table 5. The inversion of this ratio in the models was noted by Garrod (2013), and could be related to the choice of dissociation branching ratios of methanol, or to the branching ratios of the radical-addition reaction (HCO + $CH_2OH$) leading to glycolaldehyde on the grains. Whatever the cause, a reduction in the overall production of glycolaldehyde (and, by consequence, of glycolamide) in the models of around 30 times would appear to make the model ratios consistent with the observations for each of the molecules discussed here. Even with this adjustment, the modeled ratio of glycolamide to glycolaldehyde remains unchanged, suggesting that the observational upper limit obtained for glycolamide may be close to the true column density.

## 7. Conclusions

The rotational spectrum of glycolamide has been recorded up to ∼460 GHz. Precise ground state spectroscopic constants up to sextic centrifugal distortion constants are provided for the most stable conformer obtained from the fit of more than 1480 transition lines. In this work, using the heated-cell millimeter-wave spectrometer at the University of Valladolid, we were able not only to transfer this organic molecule from the solid state to the gas phase but to avoid condensation problems in the walls of the free-space cell. We used this precise set of spectroscopic parameters to search for glycolamide toward the hot core Sgr B2(N2) and the envelope of Sgr B2(N). The main conclusions of this astronomical search are the following:

1. We report a non-detection of glycolamide toward Sgr B2(N2) with ALMA with a column density upper limit of $2.4 \times 10^{16}$ cm$^{-2}$. This implies that glycolamide is at least six and five times less abundant than acetamide and glycolaldehyde, respectively, in this source. The former lower limit is only slightly higher than the column density ratio of acetaldehyde to glycolaldehyde, suggesting that the actual column density of glycolamide in Sgr B2(N2) may not be far from the derived upper limit.

2. We also report a non-detection of glycolamide toward Sgr B2(N) with the Effelsberg 100 meter radio telescope. We derive an upper limit to its beam-averaged column density of $1.2 \times 10^{14}$ cm$^{-2}$ that is similar to the column densities of glycolaldehyde and acetamide reported earlier toward the same source with the Green Bank Telescope.

3. Our astrochemical models suggest that glycolamide is a product of atomic-H abstraction from glycolaldehyde, followed by radical addition. Relative fractional abundances are close to the observed upper limit for glycolamide in Sgr B2(N2), suggesting that a slightly more sensitive search in this source could indeed yield a detection.

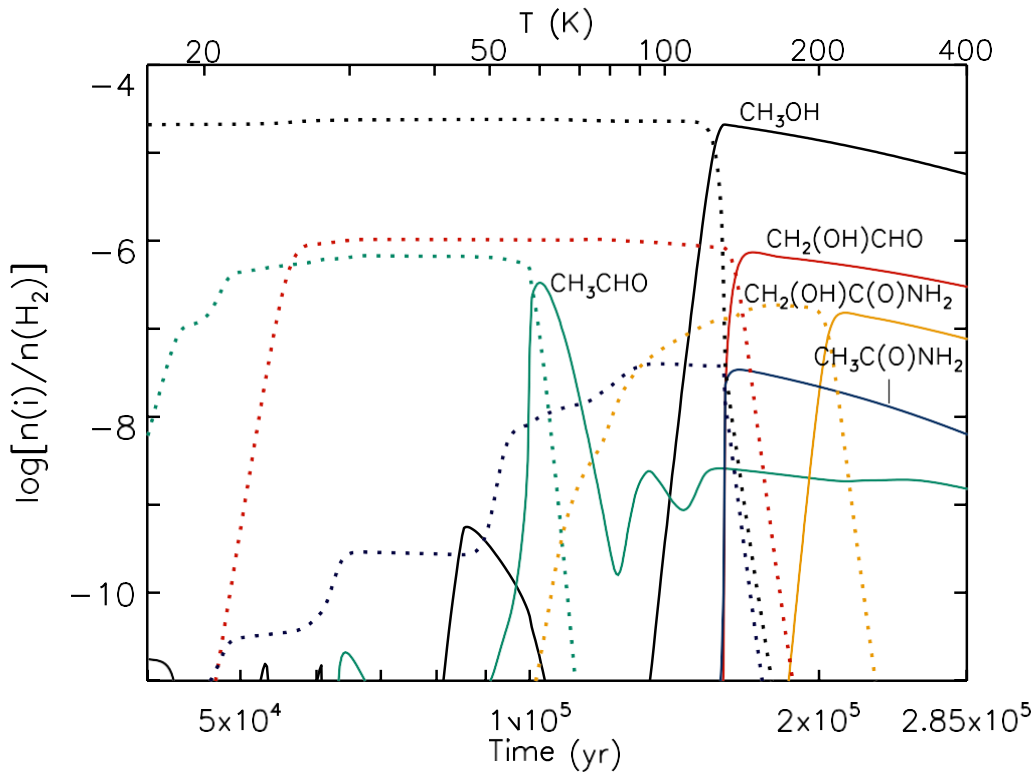

**Fig. 6.** Chemical model results for the molecules whose observational data are shown in Table 5; methanol is included for reference. Solid lines indicate gas-phase abundances with respect to H$_2$. Dashed lines of the same color indicate the solid-phase abundance of the same species, also with respect to gas-phase H$_2$. Data shown correspond to the warm-up period following free-fall collapse.

*Acknowledgements.* This research was supported by the Ministerio de Ciencia e Innovación (CTQ2013- 40717-P and CTQ2016-76393-P), Junta de Castilla y León (Grants VA175U13 and VA077U16), and the European Research Council under the European Union's Seventh Framework Programme (FP/2007–2013)/ERC-2013-SyG, Grant Agreement no. 610256 NANOCOSMOS, is gratefully acknowledged. M.S.N. acknowledges funding from the Spanish "Ministerio de Ciencia, Innovación y Universidades" under predoctoral FPU Grant (FPU17/02987). L.K. thanks The Czech Science Foundation (GACR) for financial support (grant 19-25116Y). This paper makes use of the following ALMA data: ADS/JAO.ALMA#2011.0.00017.S, ADS/JAO.ALMA#2012.1.00012.S. ALMA is a partnership of ESO (representing its member states), NSF (USA), and NINS (Japan), together with NRC (Canada), NSC and ASIAA (Taiwan), and KASI (Republic of Korea), in cooperation with the Republic of Chile. The Joint ALMA Observatory is operated by ESO, AUI/NRAO, and NAOJ. The interferometric data are available in the ALMA archive at https://almascience.eso.org/aq/. Part of this work has been carried out within the Collaborative Research Centre 956, sub-project B3, funded by the Deutsche Forschungsgemeinschaft (DFG) – project ID 184018867. We would like to thank Dr. A. Kraus, the station manager of the MPIfR Effelsberg observatory, for scheduling our observations with the 100 meter telescope on very short notice.

## Appendix A: Complementary tables

Table A.1 shows a comparison between experimental and theoretical spectroscopic constants for glycolamide and Table A.2 lists the theoretical frequencies of the lowest vibrational modes. Ta- ble A.0 lists the noise levels and continuum levels measured around all glycolamide transitions with a lower level energy in temperature units below 10 K covered by the Effelsberg observations between 4 and 8 GHz and between 12 and 17 GHz.

**Table A.1.** Experimental (exp) and theoretical (theory) spectroscopic parameters for *syn-* and *anti-* glycolamide.

| Parameters | *syn* (exp) | *anti* (exp) | *syn* (theory) | *anti* (theory) |
|---|---|---|---|---|
| A$^{(a)}$ (MHz) | 10454.2591 (20)$^{(f)}$ | 9852.6098 (20) | 10448.9 | 9932.8 |
| B (MHz) | 4041.14440 (88) | 4125.1538 (12) | 4077.1 | 4149.0 |
| C (MHz) | 2972.08215 (78) | 2967.57875 (70) | 2987.5 | 2981.2 |
| $\Delta_J$ (kHz) | 0.678 (74) | - | - | - |
| $\Delta_{JK}$ (kHz) | 9.87 (90) | - | - | - |
| $\chi_{aa}^{(b)}$ (MHz) | 2.050 (4) | 1.615 (5) | 2.1 | 1.6 |
| $\chi_{bb}$ (MHz) | 1.926 (6) | 2.005 (10) | 1.9 | 2.0 |
| $\chi_{cc}$ (MHz) | -3.976 (6) | -3.621 (10) | -4.0 | -3.6 |
| N$^{(c)}$ | 55 | 21 | - | - |
| $\sigma^{(d)}$ (kHz) | 10.8 | 6.9 | - | - |
| $\Delta$E$^{(e)}$ (cm$^{-1}$) | - | - | 0 | 218 |

**Notes.** Theoretical calculations at MP2/aug-cc-pVTZ level of theory.$^{(a)}$ A, B, and C represent the rotational constants.$^{(b)}$ $\chi_{aa}$, $\chi_{bb}$, $\chi_{cc}$ are the diagonal elements of the $^{14}$N nuclear quadrupole coupling tensor.$^{(c)}$ N is the number of measured **hyperfine components**.$^{(d)}$ $\sigma$ is the root mean square (rms) deviation of the fit.$^{(e)}$ $\Delta$E is the electronic energy relative to the global minimum, taking into account the zero-point vibrational energy (ZPE) calculated at the same level.$^{(f)}$ Standard error in parentheses in units of the last digit.

**Table A.2.** Theoretical frequencies of the lowest vibrational modes.

| Vibrational mode | Frequency$^{(a)}$ |
|---|---|
| 1 | 78.4 |
| 2 | 183.6 |
| 3 | 293.0 |
| 4 | 444.5 |
| 5 | 447.7 |
| 6 | 500.9 |
| 7 | 634.3 |
| 8 | 645.3 |
| 9 | 827.5 |
| 10 | 1042.7 |

**Notes.**$^{(a)}$ Frequencies (in cm$^{-1}$) calculated at the B2PLYPD3/aug-cc-pVTZ level of theory. They are necessary to estimate the vibrational contribution to the partition function in order to get a proper estimate of the total column density (upper limit) of the molecule.

**Appendix B: Complementary figures: Spectra**

Figures B.1–B.10 show the transitions of $CH_3CHO$, $CH_2(OH)CHO$, and some of their isotopologs or vibrationally excited states that are covered by the EMoCA survey and contribute significantly to the signal detected toward Sgr B2(N2). Figure B.11 shows examples of spectra obtained with Effelsberg toward Sgr B2(N) between 12 and 17 GHz. It illustrates the non-detection of glycolamide at these low frequencies.

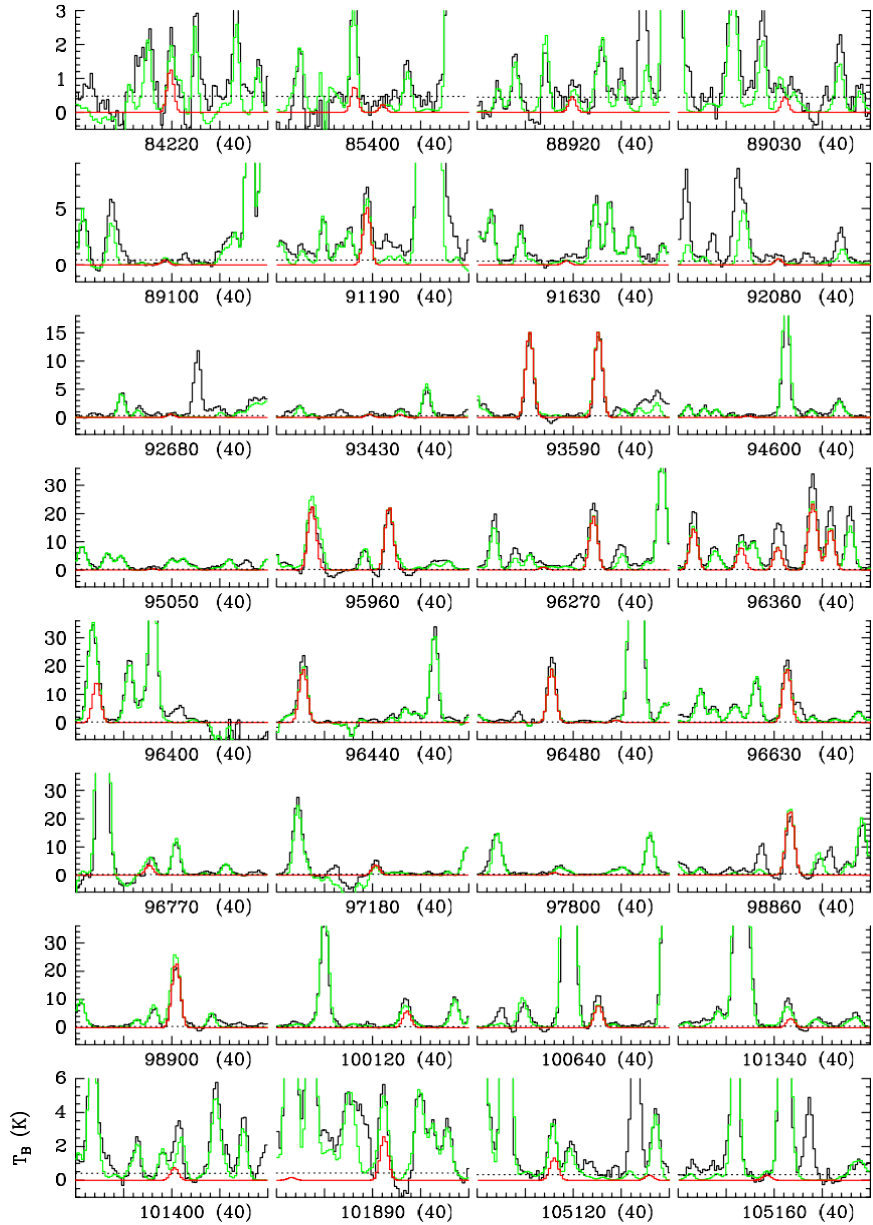

**Fig. B.1.** Transitions of CH₃CHO, r = 0 covered by our ALMA survey. The best-fit LTE synthetic spectrum of CH₃CHO, r = 0 is displayed in red and overlaid on the observed spectrum of Sgr B2(N2) shown in black. The green synthetic spectrum contains the contributions of all molecules identified in our survey so far, including the species shown in red. The central frequency and width are indicated in MHz below each panel. The y-axis is labeled in brightness temperature units (K). The dotted line indicates the 3σ noise level.

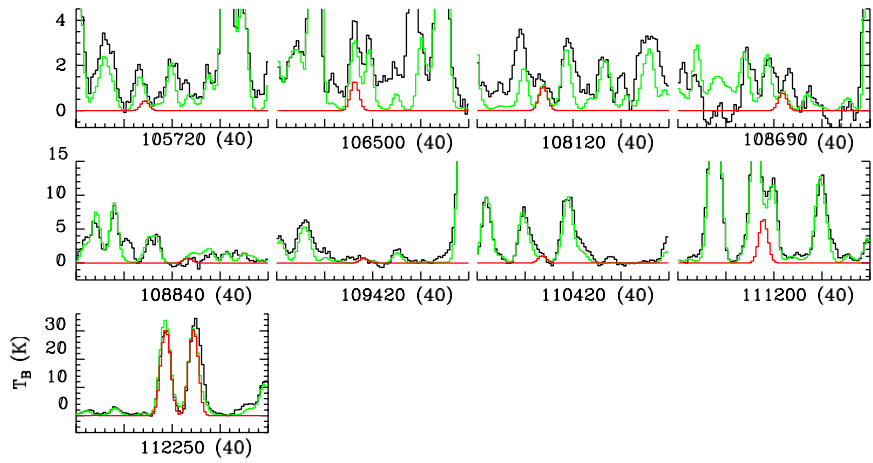

**Fig. B.1.** continued.

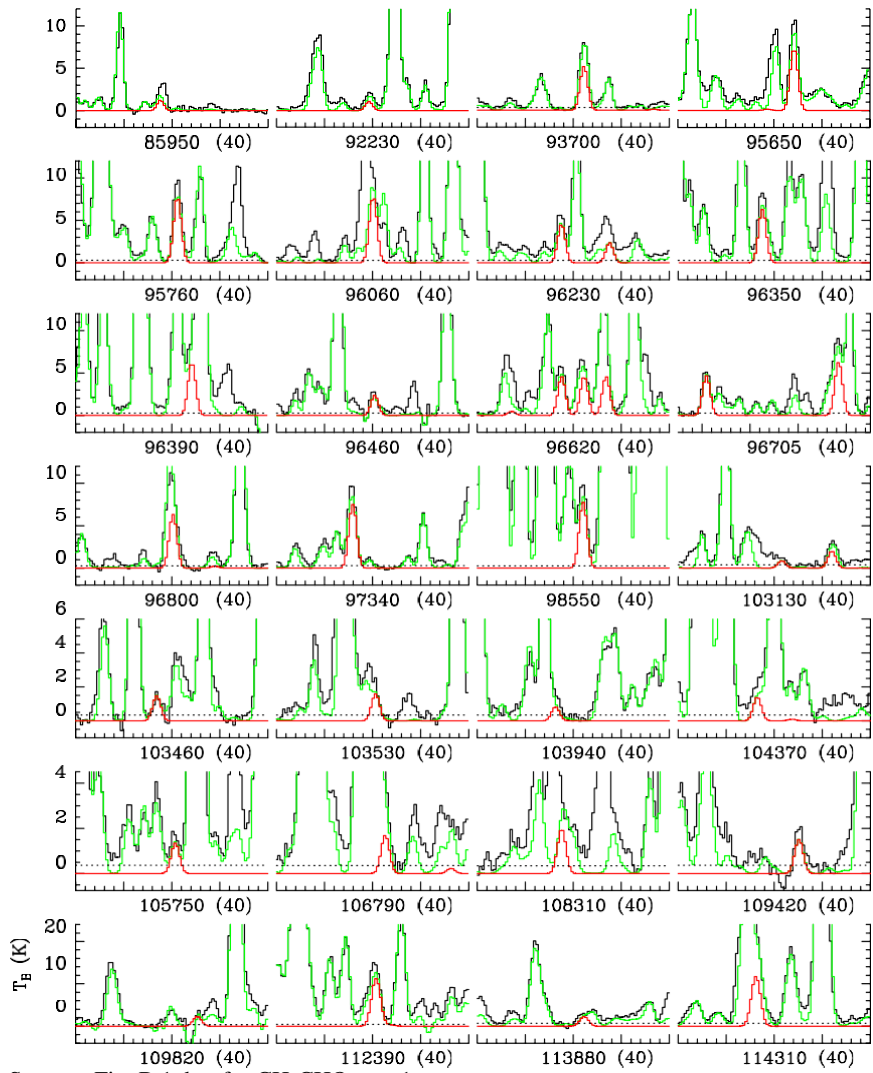

**Fig. B.2.** Same as Fig. B.1, but for CH₃CHO, $r_t = 1$.

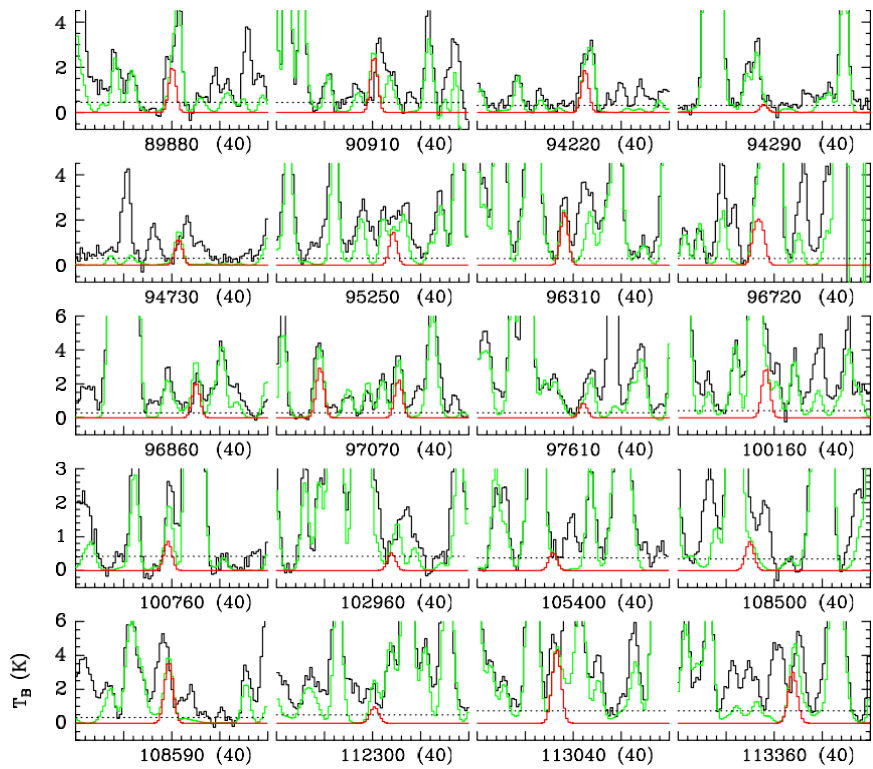

**Fig. B.3.** Same as Fig. B.1, but for CH$_3$CHO, $r_t = 2$.

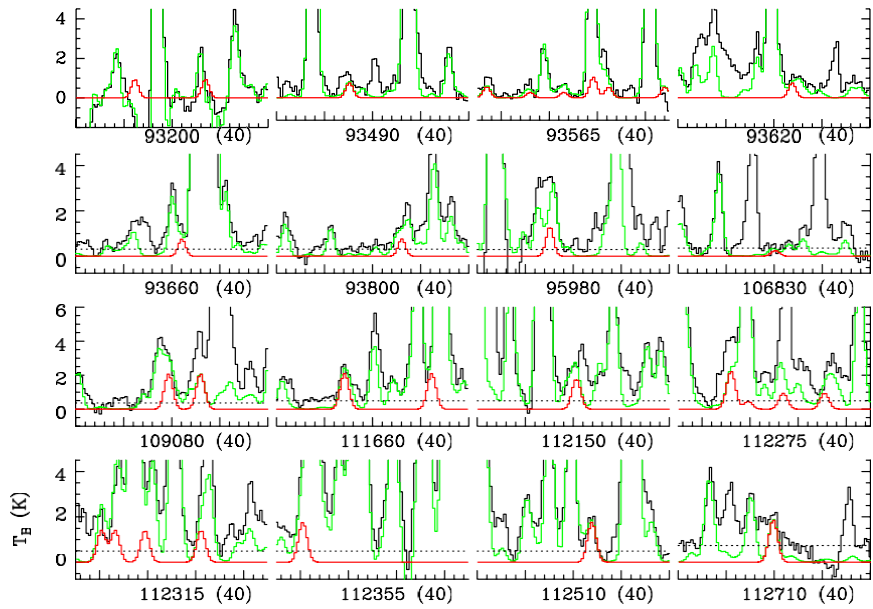

**Fig. B.4.** Same as Fig. B.1, but for $^{13}CH_3CHO$, r = 0.

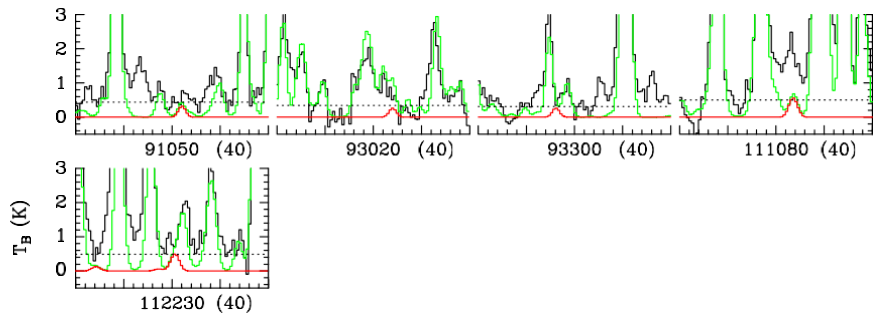

**Fig. B.5.** Same as Fig. B.1, but for $^{13}CH_3CHO$, $r_t$ = 1.

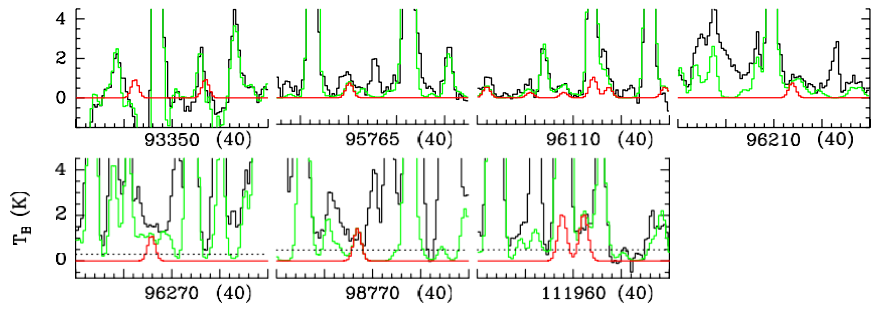

**Fig. B.6.** Same as Fig. B.1, but for CH₃¹³CHO, r = 0.

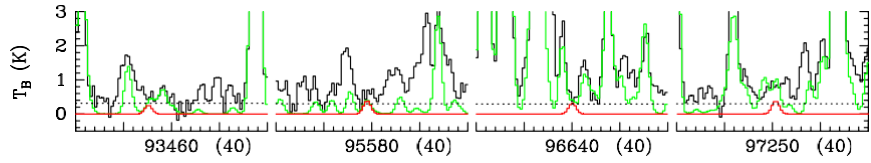

**Fig. B.7.** Same as Fig. B.1, but for CH₃¹³CHO, rₜ = 1.

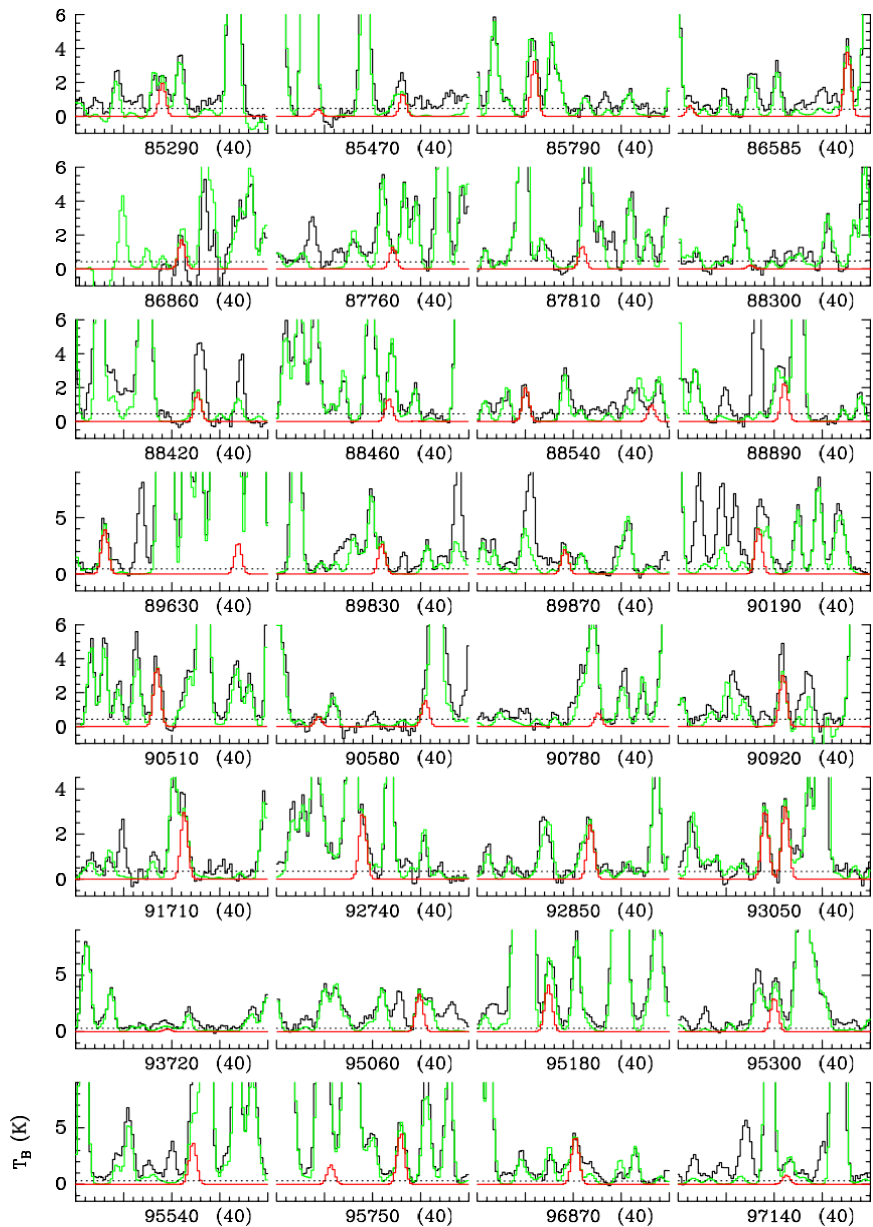

**Fig. B.8.** Same as Fig. B.1, but for CH$_2$(OH)CHO, r = 0.

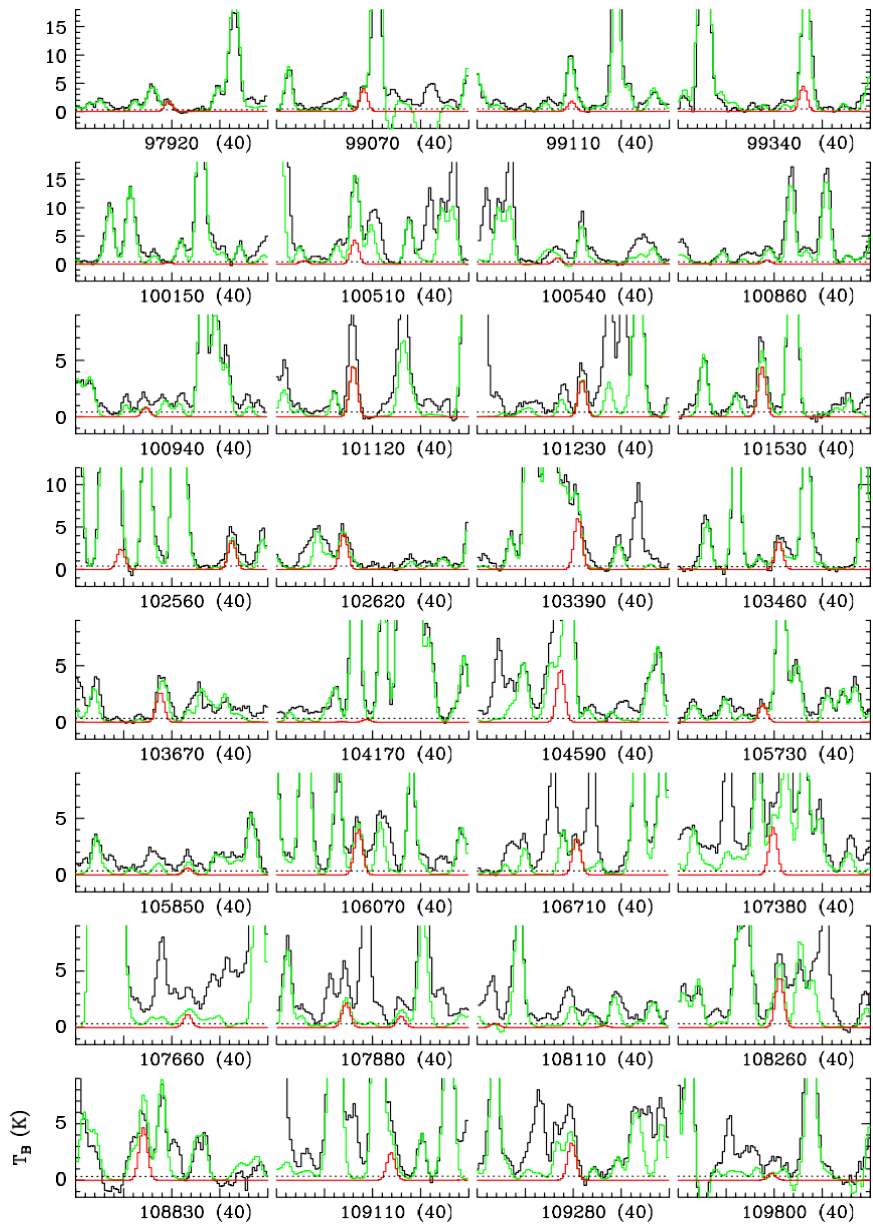

**Fig. B.8.** continued.

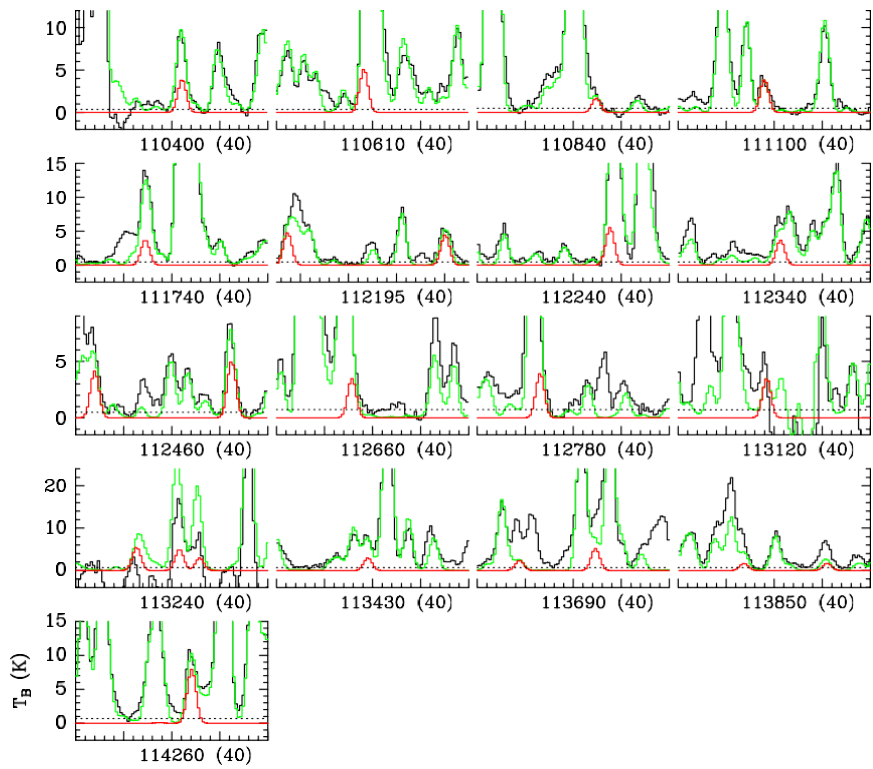

**Fig. B.8.** continued.

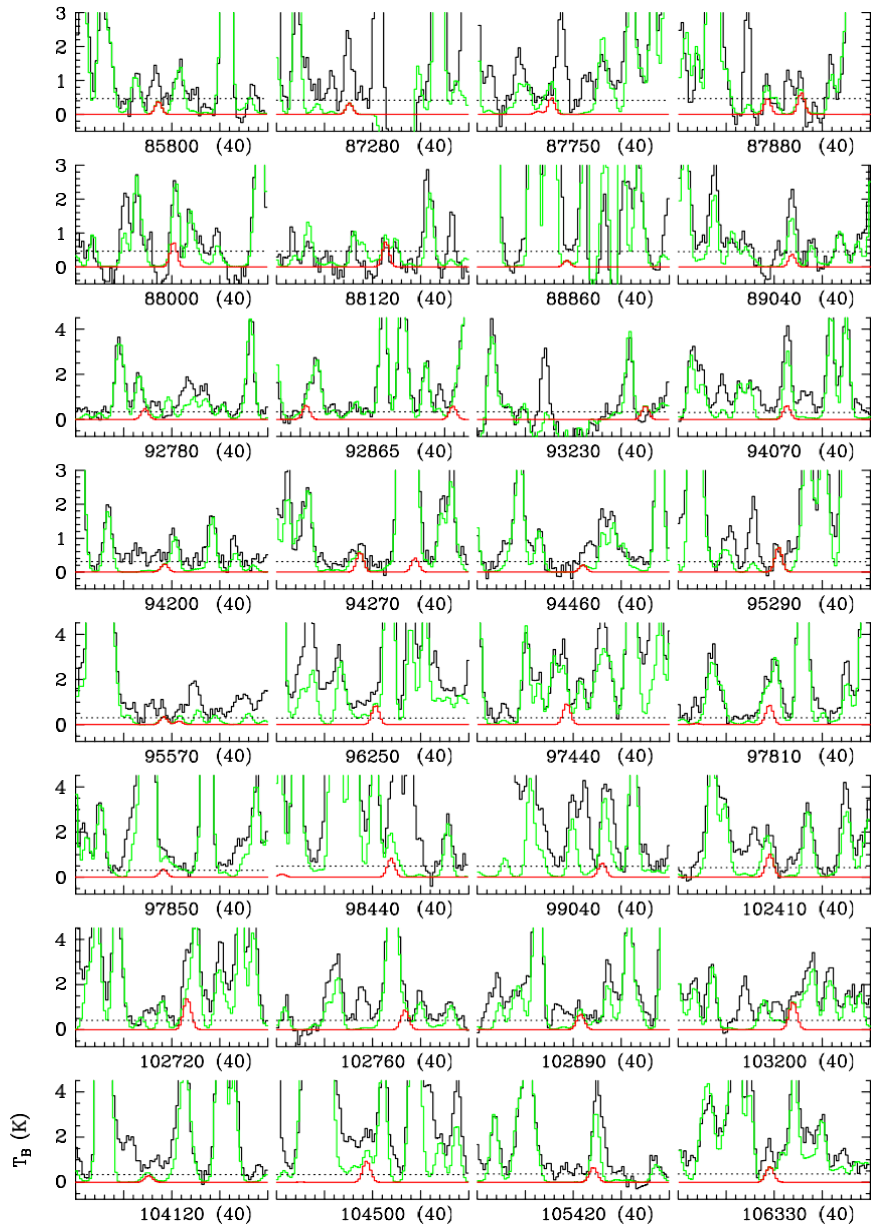

**Fig. B.9.** Same as Fig. B.1 for $CH_2(OH)CHO$, $r_1 = 1$.

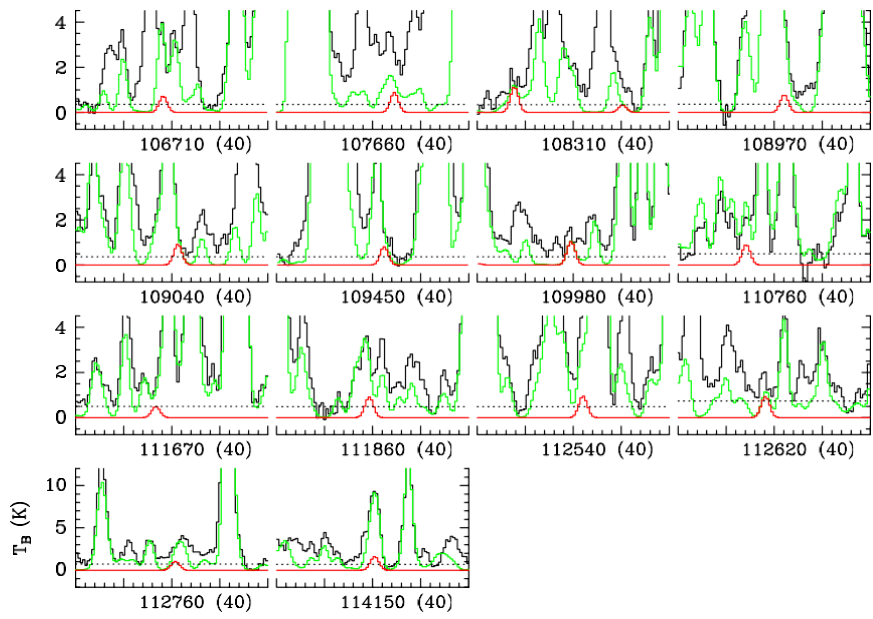

**Fig. B.9.** continued.

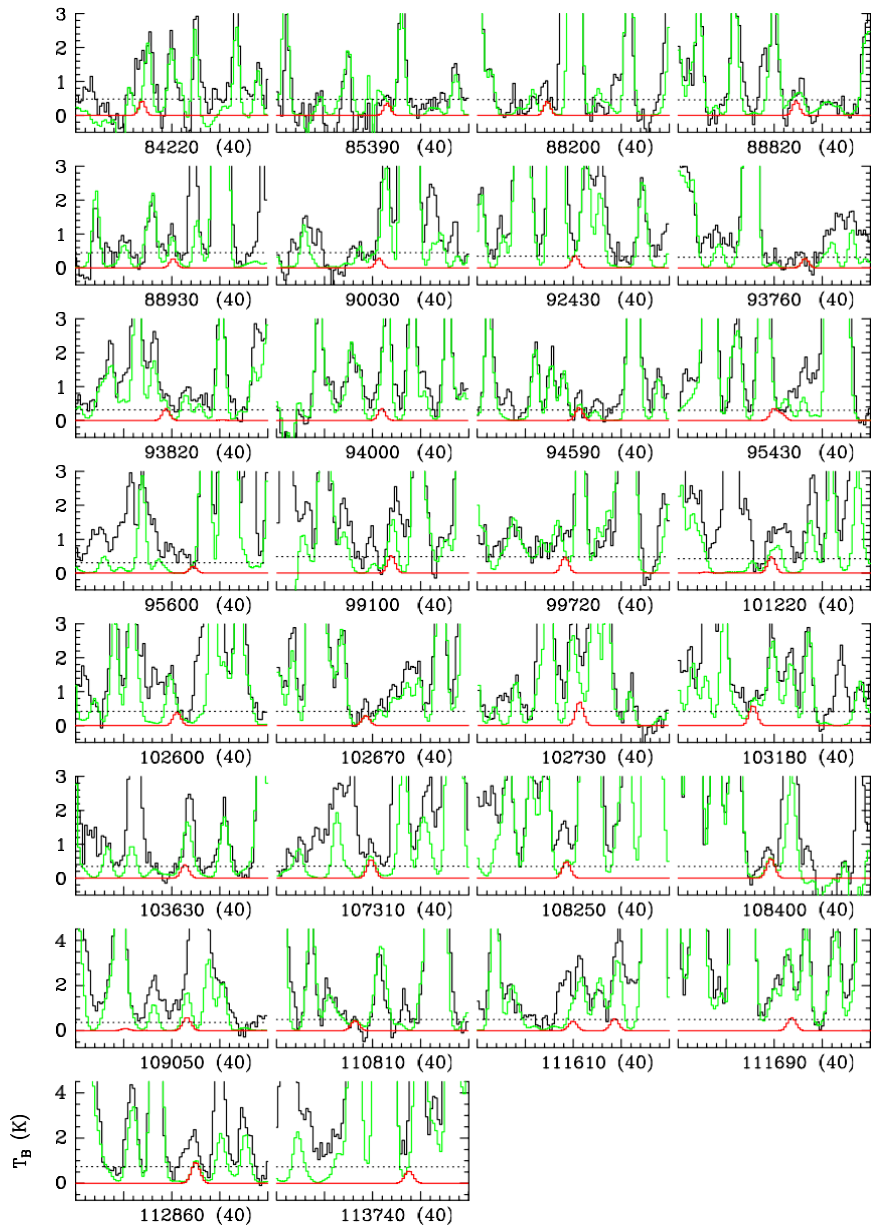

**Fig. B.10.** Same as Fig. B.1, but for CH$_2$(OH)CHO, $r_2 = 1$.

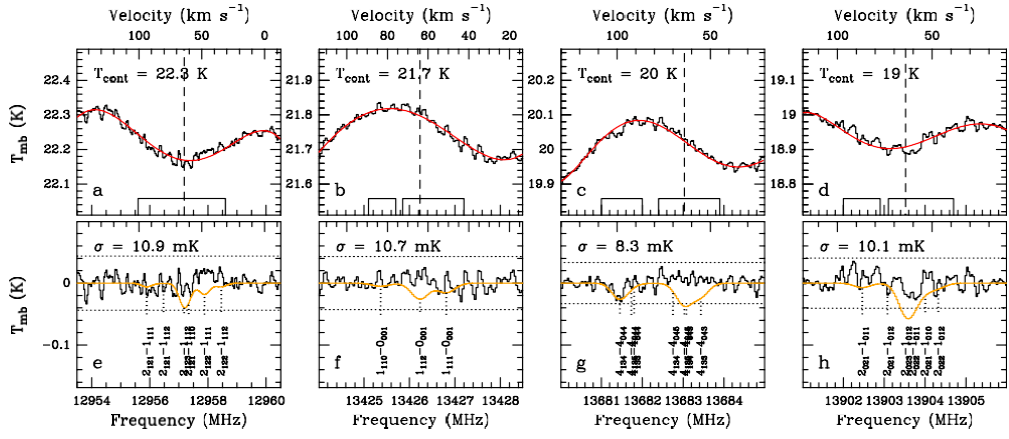

**Fig. B.11.** Same as Fig. 5, but for spectra obtained with Effelsberg between 12 and 17 GHz.

## Appendix C: Complementary figures: Population diagrams

Figures C.1 and C.2 show the population diagrams of CH$_3$CHO and CH$_2$(OH)CHO toward Sgr B2(N2).

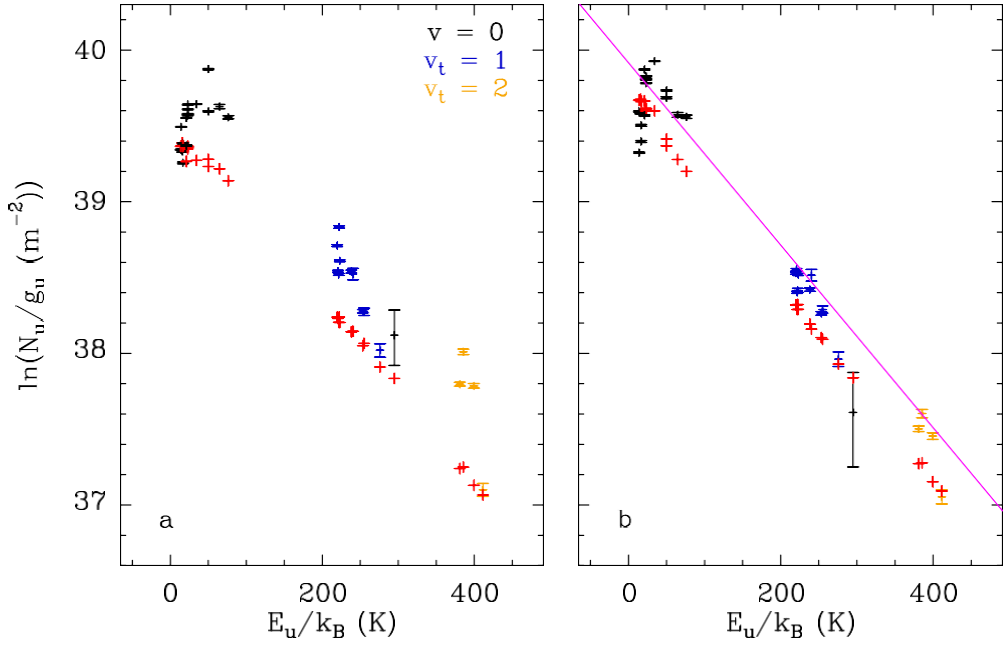

**Fig. C.1.** Population diagram of CH$_3$CHO toward Sgr B2(N2). The observed datapoints are shown in various colors (but not red) as indicated in the upper right corner of panel a while the synthetic populations are shown in red. No correction is applied in panel a. In panel b, the optical depth correction has been applied to both the observed and synthetic populations and the contamination by all other species included in the full model has been removed from the observed datapoints. The purple line is a linear fit to the observed populations (in linear-logarithmic space).

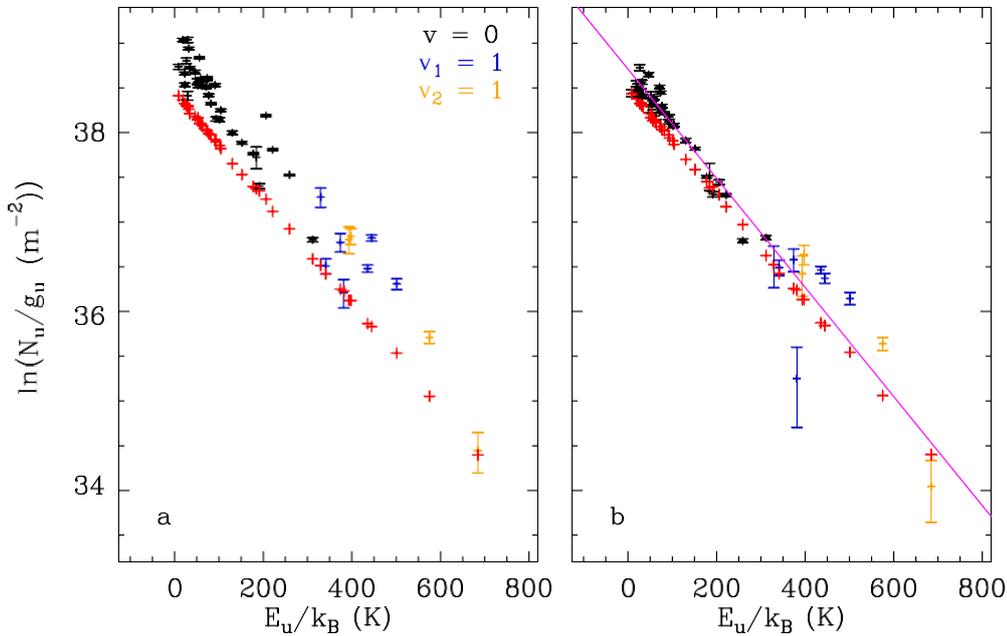

**Fig. C.2.** Same as Fig. C.1, but for CH$_2$(OH)CHO.

**Table A.0.** Spectral noise levels and continuum levels for glycolamide transitions with $E_{low}$ < 10 K observed with Effelsberg toward Sgr B2(N).

| Frequency | Transition | | | | | | | | | $E_l/k^a$ | $A_{ul}^b$ | $g_u^c$ | rms$^d$ | $T_{cont}^e$ |
|---|---|---|---|---|---|---|---|---|---|---|---|---|---|---|
| (MHz) | $J^u$ | $K_a^u$ | $K_c^u$ | $F^u$ | | $J^l$ | $K_a^l$ | $K_c^l$ | $F^l$ | (K) | ($10^{-8}$ s$^{-1}$) | | (mK) | (K) |
| 4538.854 | 7 | 1 | 7 | 8 | – | 6 | 2 | 4 | 7 | 8.694 | 0.012 | 17 | 13.4 | 55.7 |
| 5043.746 | 6 | 1 | 6 | 5 | – | 5 | 2 | 3 | 4 | 6.548 | 0.024 | 11 | 13.2 | 51.0 |
| 5044.140 | 6 | 1 | 6 | 7 | – | 5 | 2 | 3 | 6 | 6.548 | 0.024 | 15 | | |
| 5046.409 | 6 | 1 | 6 | 6 | – | 5 | 2 | 3 | 5 | 6.548 | 0.024 | 13 | | |
| 5108.735 | 3 | 3 | 1 | 4 | – | 4 | 2 | 2 | 5 | 4.779 | 0.026 | 9 | 11.3 | 50.8 |
| 5550.175 | 6 | 2 | 5 | 5 | – | 5 | 3 | 2 | 4 | 8.092 | 0.062 | 11 | 6.0 | 46.5 |
| 5550.320 | 6 | 2 | 5 | 7 | – | 5 | 3 | 2 | 6 | 8.092 | 0.064 | 15 | | |
| 5551.151 | 6 | 2 | 5 | 6 | – | 5 | 3 | 2 | 5 | 8.092 | 0.062 | 13 | | |
| 6404.012 | 3 | 1 | 2 | 3 | – | 3 | 1 | 3 | 3 | 2.196 | 0.220 | 7 | 8.8 | 44.2 |
| 6405.972 | 3 | 1 | 2 | 4 | – | 3 | 1 | 3 | 4 | 2.195 | 0.246 | 9 | | |
| 6406.658 | 3 | 1 | 2 | 2 | – | 3 | 1 | 3 | 2 | 2.195 | 0.233 | 5 | | |
| 6861.504 | 3 | 3 | 0 | 3 | – | 4 | 2 | 3 | 4 | 4.695 | 0.060 | 7 | 8.3 | 38.3 |
| 6862.721 | 3 | 3 | 0 | 4 | – | 4 | 2 | 3 | 5 | 4.695 | 0.060 | 9 | | |
| 6863.118 | 3 | 3 | 0 | 2 | – | 4 | 2 | 3 | 3 | 4.695 | 0.064 | 5 | | |
| 7001.013 | 6 | 2 | 4 | 6 | – | 6 | 2 | 5 | 6 | 8.358 | 0.315 | 13 | 7.4 | 36.0 |
| 7002.061 | 6 | 2 | 4 | 7 | – | 6 | 2 | 5 | 7 | 8.358 | 0.324 | 15 | | |
| 7002.238 | 6 | 2 | 4 | 5 | – | 6 | 2 | 5 | 5 | 8.358 | 0.321 | 11 | | |
| 7012.199 | 1 | 0 | 1 | 0 | – | 0 | 0 | 0 | 1 | 0.000 | 1.370 | 1 | 6.8 | 36.3 |
| 7013.121 | 1 | 0 | 1 | 2 | – | 0 | 0 | 0 | 1 | 0.000 | 1.371 | 5 | | |
| 7013.736 | 1 | 0 | 1 | 1 | – | 0 | 0 | 0 | 1 | 0.000 | 1.371 | 3 | | |
| 7480.660 | 1 | 1 | 0 | 1 | – | 1 | 0 | 1 | 1 | 0.337 | 0.548 | 3 | 9.8 | 37.9 |
| 7481.275 | 1 | 1 | 0 | 1 | – | 1 | 0 | 1 | 2 | 0.337 | 0.914 | 3 | | |
| 7481.853 | 1 | 1 | 0 | 2 | – | 1 | 0 | 1 | 1 | 0.337 | 0.548 | 5 | | |
| 7482.198 | 1 | 1 | 0 | 1 | – | 1 | 0 | 1 | 0 | 0.337 | 0.731 | 3 | | |
| 7482.468 | 1 | 1 | 0 | 2 | – | 1 | 0 | 1 | 2 | 0.337 | 1.646 | 5 | | |
| 7483.643 | 1 | 1 | 0 | 0 | – | 1 | 0 | 1 | 1 | 0.337 | 2.195 | 1 | | |
| 7489.391 | 2 | 0 | 2 | 1 | – | 1 | 1 | 1 | 1 | 0.644 | 0.227 | 3 | 9.6 | 37.6 |
| 7490.412 | 2 | 0 | 2 | 3 | – | 1 | 1 | 1 | 2 | 0.644 | 0.545 | 7 | | |
| 7490.632 | 2 | 0 | 2 | 2 | – | 1 | 1 | 1 | 1 | 0.644 | 0.409 | 5 | | |
| 7490.836 | 2 | 0 | 2 | 1 | – | 1 | 1 | 1 | 0 | 0.644 | 0.303 | 3 | | |
| 7491.210 | 2 | 0 | 2 | 2 | – | 1 | 1 | 1 | 2 | 0.644 | 0.136 | 5 | | |
| 12199.244 | 4 | 1 | 3 | 4 | – | 3 | 2 | 2 | 3 | 3.353 | 1.102 | 9 | 20.4 | 24.5 |
| 12199.244 | 4 | 1 | 3 | 4 | – | 3 | 2 | 2 | 4 | 3.353 | 0.074 | 9 | | |

**Table A.0.** continued.

| Frequency | | | | Transition | | | | | | $E_1/k^a$ | $A_{ul}^b$ | $g_u^c$ | rms$^d$ | $T_{cont}^e$ |
|---|---|---|---|---|---|---|---|---|---|---|---|---|---|---|
| 12199.522 | 4 | 1 | 3 | 5 | – | 3 | 2 | 2 | 4 | 3.353 | 1.175 | 11 | | |
| 12199.593 | 4 | 1 | 3 | 3 | – | 3 | 2 | 2 | 3 | 3.353 | 0.094 | 7 | | |
| 12199.593 | 4 | 1 | 3 | 3 | – | 3 | 2 | 2 | 2 | 3.353 | 1.079 | 7 | | |
| 12786.796 | 4 | 4 | 1 | 4 | – | 5 | 3 | 2 | 5 | 8.092 | 0.252 | 9 | 8.1 | 23.6 |
| 12787.652 | 4 | 4 | 1 | 5 | – | 5 | 3 | 2 | 6 | 8.092 | 0.254 | 11 | | |
| 12787.880 | 4 | 4 | 1 | 3 | – | 5 | 3 | 2 | 4 | 8.092 | 0.263 | 7 | | |
| 12795.512 | 6 | 2 | 4 | 6 | – | 5 | 3 | 3 | 5 | 8.080 | 0.940 | 13 | 9.8 | 23.3 |
| 12795.816 | 6 | 2 | 4 | 7 | – | 5 | 3 | 3 | 6 | 8.080 | 0.967 | 15 | | |
| 12795.865 | 6 | 2 | 4 | 5 | – | 5 | 3 | 3 | 4 | 8.080 | 0.935 | 11 | | |
| 12955.894 | 2 | 1 | 2 | 1 | – | 1 | 1 | 1 | 1 | 0.644 | 3.241 | 3 | 10.9 | 22.3 |
| 12956.471 | 2 | 1 | 2 | 1 | – | 1 | 1 | 1 | 2 | 0.644 | 0.216 | 3 | | |
| 12957.182 | 2 | 1 | 2 | 3 | – | 1 | 1 | 1 | 2 | 0.644 | 7.779 | 7 | | |
| 12957.338 | 2 | 1 | 2 | 1 | – | 1 | 1 | 1 | 0 | 0.644 | 4.321 | 3 | | |
| 12957.882 | 2 | 1 | 2 | 2 | – | 1 | 1 | 1 | 1 | 0.644 | 5.834 | 5 | | |
| 12958.460 | 2 | 1 | 2 | 2 | – | 1 | 1 | 1 | 2 | 0.644 | 1.945 | 5 | | |
| 13030.677 | 4 | 4 | 0 | 4 | – | 5 | 3 | 3 | 5 | 8.080 | 0.266 | 9 | 12.2 | 23.3 |
| 13031.620 | 4 | 4 | 0 | 5 | – | 5 | 3 | 3 | 6 | 8.080 | 0.268 | 11 | | |
| 13031.866 | 4 | 4 | 0 | 3 | – | 5 | 3 | 3 | 4 | 8.080 | 0.278 | 7 | | |
| 13425.365 | 1 | 1 | 1 | 0 | – | 0 | 0 | 0 | 1 | 0.000 | 8.450 | 1 | 10.7 | 21.7 |
| 13426.232 | 1 | 1 | 1 | 2 | – | 0 | 0 | 0 | 1 | 0.000 | 8.453 | 5 | | |
| 13426.810 | 1 | 1 | 1 | 1 | – | 0 | 0 | 0 | 1 | 0.000 | 8.452 | 3 | | |
| 13681.452 | 4 | 1 | 3 | 4 | – | 4 | 0 | 4 | 4 | 3.282 | 8.173 | 9 | 8.3 | 20.0 |
| 13681.730 | 4 | 1 | 3 | 5 | – | 4 | 0 | 4 | 4 | 3.282 | 0.363 | 11 | | |
| 13681.801 | 4 | 1 | 3 | 3 | – | 4 | 0 | 4 | 4 | 3.282 | 0.566 | 7 | | |
| 13682.749 | 4 | 1 | 3 | 4 | – | 4 | 0 | 4 | 5 | 3.282 | 0.443 | 9 | | |
| 13683.027 | 4 | 1 | 3 | 5 | – | 4 | 0 | 4 | 5 | 3.282 | 8.697 | 11 | | |
| 13683.083 | 4 | 1 | 3 | 4 | – | 4 | 0 | 4 | 3 | 3.282 | 0.440 | 9 | | |
| 13683.432 | 4 | 1 | 3 | 3 | – | 4 | 0 | 4 | 3 | 3.282 | 8.493 | 7 | | |
| 13902.465 | 2 | 0 | 2 | 1 | – | 1 | 0 | 1 | 1 | 0.337 | 5.315 | 3 | 10.1 | 19.0 |
| 13903.080 | 2 | 0 | 2 | 1 | – | 1 | 0 | 1 | 2 | 0.337 | 0.354 | 3 | | |
| 13903.523 | 2 | 0 | 2 | 3 | – | 1 | 0 | 1 | 2 | 0.337 | 12.760 | 7 | | |
| 13903.706 | 2 | 0 | 2 | 2 | – | 1 | 0 | 1 | 1 | 0.337 | 9.569 | 5 | | |
| 13904.002 | 2 | 0 | 2 | 1 | – | 1 | 0 | 1 | 0 | 0.337 | 7.088 | 3 | | |
| 13904.321 | 2 | 0 | 2 | 2 | – | 1 | 0 | 1 | 2 | 0.337 | 3.190 | 5 | | |
| 15090.186 | 3 | 0 | 3 | 2 | – | 2 | 1 | 2 | 2 | 1.266 | 1.107 | 5 | 30.5 | 18.5 |

**Table A.0.** continued.

| Frequency | | | | Transition | | | | | | $E_l/k^a$ | $A_{ul}^b$ | $g_u^c$ | rms$^d$ | $T_{cont}^e$ |
|---|---|---|---|---|---|---|---|---|---|---|---|---|---|---|
| 15091.579 | 3 | 0 | 3 | 3 | – | 2 | 1 | 2 | 2 | 1.266 | 6.322 | 7 | | |
| 15091.825 | 3 | 0 | 3 | 4 | – | 2 | 1 | 2 | 3 | 1.266 | 7.112 | 9 | | |
| 15092.174 | 3 | 0 | 3 | 2 | – | 2 | 1 | 2 | 1 | 1.266 | 5.974 | 5 | | |
| 15092.857 | 3 | 0 | 3 | 3 | – | 2 | 1 | 2 | 3 | 1.266 | 0.790 | 7 | | |
| 15093.987 | 2 | 1 | 1 | 1 | – | 1 | 1 | 0 | 0 | 0.696 | 6.834 | 3 | 30.5 | 18.5 |
| 15094.813 | 2 | 1 | 1 | 2 | – | 1 | 1 | 0 | 2 | 0.696 | 3.075 | 5 | | |
| 15095.432 | 2 | 1 | 1 | 3 | – | 1 | 1 | 0 | 2 | 0.696 | 12.301 | 7 | | |
| 15095.777 | 2 | 1 | 1 | 1 | – | 1 | 1 | 0 | 2 | 0.696 | 0.342 | 3 | | |
| 15096.006 | 2 | 1 | 1 | 2 | – | 1 | 1 | 0 | 1 | 0.696 | 9.225 | 5 | | |
| 15096.969 | 2 | 1 | 1 | 1 | – | 1 | 1 | 0 | 1 | 0.696 | 5.126 | 3 | | |
| 15791.081 | 5 | 1 | 4 | 5 | – | 5 | 1 | 5 | 5 | 4.961 | 1.550 | 11 | 12.6 | 14.9 |
| 15793.038 | 5 | 1 | 4 | 6 | – | 5 | 1 | 5 | 6 | 4.961 | 1.614 | 13 | | |
| 15793.436 | 5 | 1 | 4 | 4 | – | 5 | 1 | 5 | 4 | 4.961 | 1.593 | 9 | | |

**Notes.** $^{(a)}$ Energy of lower level in temperature units. $^{(b)}$ Einstein coefficient for spontaneous emission. $^{(c)}$ Degeneracy of upper energy level. $^{(d)}$ Noise level measured in baseline-removed spectrum at 38.1 kHz resolution. $^{(e)}$ Average continuum level measured around the transition or hyperfine multiplet.